\documentclass[preprintnumbers,prd,showpacs,floatfix,superscriptaddress,nofootinbib,twocolumn, letterpaper]{revtex4-1}
\usepackage{longtable}
\usepackage{natbib}
\usepackage{textcase}
\usepackage{url}
\usepackage{bm}
\usepackage{relsize}
\usepackage{amsfonts}
\usepackage{amsmath}
\usepackage{amssymb,epsf}
\usepackage{latexsym}
\usepackage{graphicx,epsfig}
\usepackage{amssymb}
\usepackage{float}
\usepackage{subfigure}
\usepackage{epstopdf}
\usepackage[colorlinks=true,citecolor=blue,linkcolor=blue,urlcolor=black]{hyperref}
\usepackage{dcolumn}
\usepackage{psfrag}
\usepackage{wrapfig}
\usepackage{makeidx}
\usepackage{epsf}
\usepackage{multirow}
\usepackage{xcolor}
\usepackage{mathtools}
\usepackage[normalem]{ulem}
\useunder{\uline}{\ul}{}

\begin{document}

\title{Gravitational baryogenesis in $f(T,L_m)$ gravity}

\author{Daniel F.P. Cruz}
\email{fc56684@alunos.fc.ul.pt}

\author{David S. Pereira}
\email{djpereira@fc.ul.pt}

\author{Francisco S.N. Lobo}
\email{fslobo@fc.ul.pt}

\affiliation{Departamento de F\'{i}sica, Faculdade de Ci\^{e}ncias da Universidade de Lisboa, Campo Grande, Edif\'{\i}cio C8, P-1749-016 Lisbon, Portugal}
\affiliation{Instituto de Astrof\'{\i}sica e Ci\^{e}ncias do Espa\c{c}o, Faculdade de
Ci\^encias da Universidade de Lisboa, Campo Grande, Edif\'{\i}cio C8,
P-1749-016 Lisbon, Portugal}

\begin{abstract}
	The observed matter--antimatter asymmetry of the Universe remains a fundamental challenge in modern physics. In this work, we explore gravitational baryogenesis within the framework of $f(T,L_m)$ gravity, where the gravitational Lagrangian depends on both the torsion scalar $T$ and the matter Lagrangian $L_m$. We consider three representative models and examine their ability to generate the observed baryon-to-entropy ratio. Our analysis shows that couplings involving both torsion and the matter Lagrangian, $\partial_\mu(-T-\frac{L_m}{L_0})$, can successfully account for the baryon asymmetry for decoupling temperatures in the range $10^{12}$--$10^{14}\,\text{GeV}$, while remaining consistent with small deviations from General Relativity. These results highlight the capacity of $f(T,L_m)$ gravity to provide novel mechanisms for baryogenesis, demonstrating that the interplay between torsion and matter-sector contributions can naturally generate the observed asymmetry. The framework also remains compatible with late-time cosmological evolution, offering a unified setting for early- and late-time dynamics.
\end{abstract}
\date{\today}
\maketitle

\section{Introduction}

The observed asymmetry between matter and antimatter remains one of the most persistent and fundamental problems in modern cosmology~\cite{Dimopoulos:1978kv,wilczek1980cosmic,Dine:2003ax,Cline:2006ts}. Despite the fact that the Standard Model of particle physics provides a highly successful description of microphysical processes, it does not adequately explain why the Universe today is overwhelmingly dominated by matter. This imbalance is customarily quantified in terms of the baryon-to-entropy ratio, defined as
\begin{equation}
	\frac{n_b}{s} \equiv \frac{n_B - n_{\Bar{B}}}{s}\,,
\end{equation}
where $n_B$ and $n_{\Bar{B}}$ denote the number densities of baryons and anti-baryons, respectively, and $s$ represents the entropy density of the relativistic radiation bath. In a perfectly symmetric Universe, one would expect $n_B = n_{\Bar{B}}$, and thus $n_b/s = 0$. However, cosmological observations clearly indicate a nonzero value. 

More concretely, precise measurements from the anisotropies of the Cosmic Microwave Background (CMB), combined with constraints from Big Bang Nucleosynthesis (BBN), establish that the baryon-to-entropy ratio is of order $10^{-10}$. Current observational data constrain this quantity to be~\cite{wilczek1980cosmic,Burles:2000ju,WMAP:2003ivt,Burles:2000zk,Planck:2018vyg,Fields:2019pfx,WMAP:2003ogi,ParticleDataGroup:2020ssz}
\begin{equation}\label{observational}
	\frac{n_b}{s} \simeq (8.8 \pm 0.6)\times 10^{-11} .
\end{equation}
This extremely small yet nonvanishing ratio encodes one of the key empirical clues about the physics of the early Universe, and any successful theory of baryogenesis must account for its observed value.

Since the pioneering work of Sakharov~\cite{Sakharov}, it has been understood that there are three key ingredients that combined can dynamically generate a baryon asymmetry in the early Universe: the violation of baryon number, the violation of C and CP symmetries, and a departure from thermal equilibrium. These criteria, now known as the Sakharov conditions, have guided theoretical model building for decades, and a wide range of mechanisms have been proposed to account for the observed matter--antimatter imbalance~\cite{Bodeker:2020ghk,Pereira:2023xiw,Garbrecht:2018mrp,Allahverdi:2012ju,Morrissey:2012db,Riotto:1999yt}. 

Among these proposals, \emph{gravitational baryogenesis}~\cite{Davoudiasl:2004gf} has emerged as a particularly intriguing scenario. The central idea relies on the presence of a higher-dimensional operator in the effective Lagrangian of the form
\begin{equation}
	\mathcal{L}_{\text{GB}} = \frac{1}{M_*^2}\,\partial_\mu R\, J^\mu_B\,, \label{L_int}
\end{equation}
where $M_*$ denotes a characteristic cutoff scale, $R$ is the Ricci scalar, and $J^\mu_B$ is the baryonic current. This interaction explicitly violates CPT symmetry and can naturally arise in the low-energy effective field theory of quantum gravity or in supergravity frameworks if the cutoff is near the Planck scale, $M_* \sim M_{\text{Pl}}$, with $M_{\text{Pl}} \simeq 2.4 \times 10^{18}\,\text{GeV}$ being the reduced Planck mass~\cite{Davoudiasl:2004gf}. 

The mechanism operates by inducing an effective chemical potential for baryon number through the coupling of the time-varying curvature scalar to the baryonic current. In contrast to conventional baryogenesis models, which rely on all three Sakharov conditions, gravitational baryogenesis only requires the existence of baryon number violating interactions. The effective CPT-violating term biases the thermal equilibrium in favor of baryons over anti-baryons, thereby enabling the generation of the observed asymmetry without the necessity of a departure from equilibrium. This provides a minimal and conceptually distinct realization of baryogenesis that relaxes some of the traditional Sakharov requirements.

This mechanism has attracted substantial interest, particularly within the framework of modified theories of gravity~\cite{Pereira:2025flo,Mojahed:2024yus,Bhattacharjee:2020wbh,Sahoo:2019pat,Li:2004hh,Lambiase:2006dq,Odintsov:2016hgc,Bhattacharjee:2021jwm,Jaybhaye:2023lgr,Baffou:2018hpe,Oikonomou:2016jjh,Bhattacharjee:2020jfk,Mishra:2023khd,Pereira:2024ddu}, where deviations from Einstein’s General Relativity (GR) can substantially modify the dynamics of the early Universe. In such frameworks, the gravitational sector naturally acquires additional degrees of freedom or higher-order corrections, which in turn give rise to new couplings between geometry and matter fields. These modifications can enhance the efficiency of baryogenesis by effectively amplifying the CPT-violating interaction in Eq.~\eqref{L_int}, thereby broadening the theoretical landscape in which the observed baryon asymmetry might be explained.

Within the setting of teleparallel gravity, for instance, gravitational baryogenesis has been investigated in the context of $f(T)$ gravity~\cite{Oikonomou:2016jjh}, where $T$ denotes the torsion scalar, as well as in $f(T,\mathcal{T})$ gravity~\cite{Mishra:2023khd}, where $\mathcal{T}$ represents the trace of the energy--momentum tensor. In both cases, new interaction terms involving $T$ or $f(T,\mathcal{T})$ were introduced to generalize the baryogenesis mechanism beyond the standard curvature-based formulation. These extensions illustrate how alternative theories of gravity can provide novel pathways for realizing the observed baryon asymmetry, while simultaneously offering phenomenological probes of the underlying gravitational dynamics.

Beyond curvature- and torsion-based formulations, gravity can also be described in terms of the nonmetricity scalar $Q$ within the so-called symmetric teleparallel framework. Together, the curvature, torsion and nonmetricity formulations form the so-called ``geometrical trinity of gravity''~\cite{BeltranJimenez:2019esp}. In this language, GR can be reformulated equivalently as a purely nonmetricity theory, which in turn motivates generalized nonmetricity theories such as $f(Q)$ gravity. A comprehensive overview of these developments and their cosmological applications can be found in the recent review on $f(Q)$ gravity~\cite{Heisenberg:2023lru}. Building on this, further generalizations including explicit matter couplings, such as $f(Q,T)$~\cite{Xu:2019sbp} and $f(Q,L_m)$~\cite{Hazarika:2024alm} gravity, have been proposed. These theories exhibit a rich phenomenology at both early and late times and have been explored, for instance, in the context of dynamical reconstruction of $\Lambda$CDM~\cite{Kaczmarek:2024quk} and in novel baryogenesis scenarios in $f(Q,L_m)$ gravity~\cite{Samaddar:2024qno}. Our present analysis can then be viewed as the torsion-based analogue of these nonmetricity models, in which the matter sector couples non-minimally to torsion rather than to nonmetricity.

In this work, we investigate the $f(T,L_m)$ theory, an extension of $f(T)$ gravity, and explore its implications for late-time cosmology as well as within the framework of gravitational baryogenesis. Analogously to the generalization from $f(R)$ to $f(R,L_m)$ gravity---which can be regarded as the maximal extension of the Hilbert--Einstein action~\cite{fRLm}---the formulation $f(T,L_m)$ introduces a maximal coupling between the torsion scalar $T$ and the matter Lagrangian $L_m$. This extended framework enables a richer set of dynamical interactions between matter and geometry, thereby enlarging the space of viable cosmological scenarios. 

Motivated by the successes of $f(R,L_m)$ theories in addressing a wide range of gravitational and cosmological phenomena~\cite{Jaybhaye:2023lgr,fRLm,Jaybhaye:2022gxq,Harko:2014gwa,Harko:2013yb}, the extension to $f(T,L_m)$ gravity offers a natural framework to incorporate more intricate matter-sector contributions into teleparallel gravity. By introducing a direct dependence on the matter Lagrangian $L_m$ within the torsion gravitational sector, this formulation enriches the possible interactions between matter and geometry. Such an extended structure not only broadens the landscape of modified gravity theories but also opens the possibility of uncovering new mechanisms for baryogenesis. In particular, it provides novel pathways for explaining the observed matter--antimatter asymmetry of the Universe~\cite{Pereira:2025flo}.

In choosing specific $f(T,L_m)$ models for our analysis, our aim is not to propose an exhaustive classification, but rather to work with phenomenologically well-motivated and representative examples. Our work is, to the best of our knowledge, the first systematic study of gravitational baryogenesis in $f(T,L_m)$ gravity. It complements previous analyses of baryogenesis in $f(R,L_m)$, $f(T,\mathcal{T})$, $f(Q,T)$, $f(Q,L_m)$ and related theories~\cite{Bhattacharjee:2020wbh,Baffou:2018hpe,Jaybhaye:2023lgr,Azhar:2020coz,Jawad:2023poj,Alruwaili:2025mrc,Usman:2024cya,Samaddar:2024qno}, as well as the general review~\cite{Pereira:2023xiw}.
The model $f = B T L_m$ is the simplest non-minimal torsion--matter coupling and can be viewed as the $f(T,L_m)$ analogue of the linear $f(R,L_m)$ ansatz, recovering the teleparallel equivalent of GR in the limit $B \to 0$. The models $f = B(1-e^{-p\sqrt{|T|}})L_m$ and $f = B(1-e^{-pT})L_m$ are torsion--matter counterparts of the widely studied Linder-type and exponential $f(T)$ models~\cite{Linder,exponential}, which are known to provide viable late-time cosmologies. In all cases, the additional terms can be interpreted as effective dark fluids arising from torsion--matter couplings, with $B$ controlling the strength of the coupling and $p$ introducing a characteristic torsional scale. These new terms allow us to tackle some of the problems that gravitational baryogenesis has in GR. For example, in a pure radiation dominated Universe the Ricci scalar is zero, implying that no asymmetry is produced. Furthermore, by only using a perfect fluid plus GR one arrives at scenarios that produce either an insufficient or excessive asymmetry. The new terms $f(T,L_m)$ may then help resolve these inherent issues.

This paper is organized as follows. In Sec.~\ref{sec:f(T)}, we introduce the formalism of torsion-based theories of gravity, while in Sec.~\ref{sec:fTlm} we present the theoretical framework of $f(T,L_m)$ gravity. The implications of this framework in the context of gravitational baryogenesis are explored in Sec.~\ref{sec:GB fTlm}. Finally, in Sec.~\ref{sec:conclusion}, we provide a summary of our findings together with a discussion of their broader significance.

\section{Teleparallel formalism}\label{sec:f(T)}

Teleparallelism, like GR, is formulated on a general spacetime, namely a four-dimensional differentiable manifold $\mathcal{M}$, whose tangent space at each point is a Minkowski spacetime~\cite{Pereira,f(T)_and_cosmology}. In teleparallel gravity, the fundamental dynamical variables are the tetrad fields $e^{\ \mu}_{A}$, which provide an orthonormal basis for the tangent space at each point of the manifold. Greek indices $(\mu, \nu, \dots)$ are used to label the coordinates of the spacetime manifold, whereas Latin indices $(A, B, \dots)$ denote coordinates in the tangent space, with both sets of indices running from $0$ to $3$. The spacetime metric is then constructed from the tetrads as
\begin{equation}
    g_{\mu\nu}=\eta_{AB}e^{A}_{\ \mu} e^{B}_{\ \nu}\, ,
\end{equation}
where $\eta_{AB} = {\rm diag}(1, -1, -1, -1)$ is the metric tensor of
Minkowski spacetime in Cartesian coordinates.

In teleparallel gravity, the geometry of spacetime is characterized by torsion rather than curvature. Consequently, the theory employs a connection distinct from the Levi-Civita connection used in curvature-based gravity. Specifically, one adopts the curvature-free Weitzenböck connection~\cite{Pereira}, defined as
\begin{equation}
	\Gamma^\lambda_{\ \mu\nu} \equiv e_A^{\ \lambda} \partial_\nu e^A_{\ \mu}\,.
\end{equation}
The torsion tensor associated with this connection is then defined by
\begin{equation}
    T^\rho_{\mu\nu}\equiv\Gamma^\rho_{\mu\nu}-\Gamma^\rho_{\nu\mu}=e_A^{\ \rho}(\partial_\mu e^A_{\ \nu}-\partial_\nu e^A_{\ \mu})\, .
    \label{torsion}
\end{equation}

Analogous to the Ricci scalar $R$, the Lagrangian of teleparallel theory of gravity is the torsion scalar $T$
\begin{eqnarray}
    T &\equiv & T^\rho_{\ \mu\nu}S_\rho^{\ \mu\nu}
    	\nonumber \\
    &=& \frac{1}{4}T^{\rho\mu\nu}T_{\rho\mu\nu}+\frac{1}{2}T^{\rho\mu\nu}T_{\nu\mu\rho}-T_{\rho\mu}^{\ \ \ \rho}T^{\nu\mu}_{\ \ \ \nu}\,,
    \label{Torsion Scalar}
\end{eqnarray}
where $S_\rho^{\ \mu\nu}$ is the superpotential, defined as
\begin{equation}
    S_\rho^{\ \mu\nu}=\frac{1}{2}\left(K^{\mu\nu}_{\ \ \ \rho}+\delta^\mu_\rho T^{\alpha\nu}_{\ \ \ \alpha}-\delta^\nu_\rho T^{\alpha\mu}_{\ \ \ \alpha} \right)\,,
\end{equation}
and the cotorsion tensor $K^{\mu\nu}_{\ \ \ \rho}$ is given by
\begin{equation}
    K^{\mu\nu}_{\ \ \ \rho}=-\frac{1}{2}(T^{\mu\nu}_{\ \ \ \rho}-T^{\nu\mu}_{\ \ \ \rho}-T_\rho^{\ \mu\nu})\,.
\end{equation}

With the previous formalism defined, the action for $f(T)$ gravity~\cite{Oikonomou:2016jjh} is given by
\begin{equation}
    S=\frac{1}{2\kappa^2}\int e \,dx^4[T+f(T)]+\int e \, dx^4L_m\,,
\end{equation}
where $e={\rm det}(e^A_{\ \mu})=\sqrt{-g}$ with $g$ being the metric determinant,  $\kappa^2 = (8\pi G)^2=M_{\text{Pl}}^{-2}$, and $L_m$ is the matter field Lagrangian.

By varying the action with respect to the tetrad fields, we obtain the corresponding field equations
\begin{eqnarray}
        &&\left[e^{-1}\partial_\mu(ee_A^{\ \rho}S_\rho^{\ \mu\nu})-e_A^{\ \lambda}T^{\rho}_{\ \mu\lambda}S_\rho^{\ \nu\mu}\right](1+F_T)
        	\nonumber \\
        && \hspace{-0.5cm} +e_A^{\ \rho}S_\rho^{\ \mu\nu}F_{TT}\partial_\mu T
        +e_A^{\ \nu}\left(\frac{T+f}{4}\right)=4\pi G e_A^{\ \rho}\overset{em}{T}{}_\rho^{\ \nu}\,,
        \label{f(T)_fieldequations}
\end{eqnarray}
where $\overset{em}{T}{}_\rho^{\ \nu}$ is the energy momentum tensor of a perfect fluid defined as $\frac{\delta eLm}{\delta e^A_{\ \nu}}=-ee_A^{\ \rho}T_\rho^{\ \nu}$ \cite{Pereira}, and $F_T=\partial f/\partial T$, $F_{TT}=\partial^2 f/\partial T^2$.

Consider a flat FLRW universe with the following metric
\begin{equation}
    ds^2=dt^2-a^2(t)\delta_{ij}dx^idx^j\,,
\end{equation}
where $a(t)$ is the scale factor, which arises from the diagonal tetrad
\begin{equation}
    e_A^{\ \nu}=(1,a,a,a)\,.
    \label{tetrad}
\end{equation}
Inserting the tetrad (\ref{tetrad}) into the $f(T)$ field equations (\ref{f(T)_fieldequations}) we obtain the modified Friedman equations
\begin{equation}
    H^2=\frac{8\pi G}{3}\rho-2H^2F_T-\frac{f}{6}\, ,
\end{equation}
\begin{equation}
   \Dot{H}=-4\pi G(\rho+P)-\Dot{H}(F_T-12H^2F_{TT}) \,,
\end{equation}
respectively, where $H = \dot{a}/a$ denotes the Hubble parameter, with the overdot representing differentiation with respect to cosmic time $t$. The effective energy density and pressure of the matter content are denoted by $\rho$ and $P$, respectively. Using Eq.~(\ref{tetrad}) with Eqs.~(\ref{torsion}) and (\ref{Torsion Scalar}), the torsion scalar corresponding to the FLRW metric can be expressed as
\begin{equation}
    T=-6H^2\,.
    \label{T}
\end{equation}
This property will play an important role in the calculations that follow.

Naturally, other extensions have been proposed, such as $f(T,\mathcal{T})$ gravity~\cite{f(TT)}, mentioned above, and models with a non-minimal torsion–matter coupling~\cite{Harko:2014sja}. In the following section, we focus on $f(T,L_m)$ gravity, which can be regarded as a natural generalization of the non-minimal torsion–matter coupling framework~\cite{Harko:2014sja}.

\section{$f(T,L_m)$ Gravity}\label{sec:fTlm}

In this work we generalize the torsion scalar in the action to a more general function $T\to T+f(T,L_m)$ where $L_m$ is the matter field Lagrangian
\begin{equation}
    S=\frac{1}{2\kappa^2}\int e\,dx^4\ [T+f(T,L_m)]+\int e\,dx^4 \ L_m\, .
    \label{action}
\end{equation}
and we assume that the Lagrangian depends solely on the tetrads $e^A_{\ \mu}$ and not on their derivatives. Varying the action~(\ref{action}) with respect to the tetrads then yields the corresponding field equations:
\begin{eqnarray}
        &&\hspace{-0.35cm}\left[e^{-1}\partial_\mu(ee_A^{\ \rho}S_\rho^{\ \mu\nu})-e_A^{\ \lambda}T^{\rho}_{\ \mu\lambda}S_\rho^{\ \nu\mu}\right](1+F_T)
        +e_A^{\ \rho}S_\rho^{\ \mu\nu}
        	\nonumber \\
        && \quad \times \left(F_{TT}\partial_\mu T+F_{TL}\partial_\mu L_m \right)-\frac{1}{4}F_L\left(e_A^{\ \rho}\overset{em}{T}{}_\rho^{\ \nu}+e_A^{\ \nu}L_m\right)
        	\nonumber \\
        && \quad +e_A^{\ \nu}\left(\frac{T+f}{4}\right)=4\pi G e_A^{\ \rho}\overset{em}{T}{}_\rho^{\ \nu}\,.
    \label{field equations}
\end{eqnarray}

Inserting the tetrad (\ref{tetrad}) into the field equations (\ref{field equations}), we obtain the modified Friedmann equations, given by
\begin{equation}
    H^2=\frac{8\pi G}{3}\rho-2H^2F_T+\frac{1}{6}F_L\left(\rho+L_m\right)-\frac{f}{6}\, ,
    \label{first Friedmann}
\end{equation}
\begin{eqnarray}
        \Dot{H}=-4\pi G(\rho+P)-\Dot{H}(F_T-12H^2F_{TT})
        	\nonumber \\
        -HF_{TL}\partial_tL_m-\frac{1}{2}F_L\left(P+L_m\right)\,,
\end{eqnarray}
respectively. These can be written as
\begin{equation}
    H^2=\frac{8\pi G}{3}\left(\rho+\rho_{MG}\right)\label{Hubble}\, ,
\end{equation}
\begin{equation}
    2\Dot{H}+3H^2=-8\pi G \left(P+P_{MG}\right)\, ,
\end{equation}
where
\begin{equation}
    \rho_{MG}=\frac{3}{8\pi G}\left[-2H^2F_T+\frac{1}{6}F_L\left(\rho+L_m\right)-\frac{f}{6}\right]\label{rho_DE}\, ,
\end{equation}
and
\begin{eqnarray}
        P_{MG}=-\rho_{MG}+\frac{1}{8\pi G}\Big[2\Dot{H}(F_T-12H^2F_{TT})
        	\nonumber \\
        +2H\partial_t(L_m)F_{TL}+F_L(P+L_m)\Big]\, .
\end{eqnarray}

Since gravitational baryogenesis occurs during the radiation-dominated era, the total energy density is dominated by radiation, i.e., $\rho = \rho_m + \rho_r \approx \rho_r$. Assuming that the contributions from modified gravity (MG) constitute only a small correction to GR, the energy density associated with MG must be much smaller than the radiation energy density
\begin{equation}
    \left|\frac{\rho_{MG}}{\rho}\right| \ll 1\, .
\end{equation}\label{Constrain1}

Thus, in this work, we consider the case where the contribution of modified gravity is small, taking $|\rho_{\rm MG}/\rho | \lesssim 0.10.$
Using Eq.~\eqref{rho_DE}, we then obtain
\begin{equation}
    \left|\frac{\rho_{MG}}{\rho_r}\right|=\left|\frac{3M_{\text{Pl}}^2\left[-2H^2F_T+\frac{1}{6}F_L\left(\rho_r+L_m\right)-\frac{f}{6}\right]}{\rho_r}\right|\, .
\end{equation}

\section{Gravitational baryogenesis in $f(T,L_m)$ gravity}\label{sec:GB fTlm}

In the context of $f(T,L_m)$ gravity we consider three new interaction terms:
\begin{equation}\label{eq:-T term}
    \mathcal{L}_\text{int} = \frac{\epsilon}{M^2_*}(\partial_\mu(-{T}))J^\mu\,,
\end{equation}
\begin{equation}\label{CP-Violating}
    \mathcal{L}_\text{int} = \frac{\epsilon}{M^2_*}\left[\partial_\mu\left(-{T}-\frac{L_m}{L_0}\right)\right]J^\mu\,.
\end{equation}
and
\begin{equation}
    \mathcal{L}_{int}=\frac{\epsilon}{M^2_*}\big[\partial_\mu(f(T,L_m))\big]J^\mu
    \label{General_CP}\, ,
\end{equation}
where  $ L_0 $ is introduced to ensure dimensional consistency, with $ [L_0] = \text{GeV}^2 $. The interaction term \eqref{CP-Violating} consists of two terms that provide distinct contributions: the term $ \frac{1}{M_*^2} \partial_\mu(-T) J^\mu $, associated with a cutoff energy scale $ \Lambda_1 = M_* $, and the term $ \frac{1}{M_*^2} \partial_\mu\left(-\frac{L_m}{L_0}\right) J^\mu $, which corresponds to a cutoff energy scale $ \Lambda_2 = \left(M_*^2 L_0\right)^{1/4} $. In this work, we adopt $ L_0 = M_*^2 $ to ensure that $ \Lambda_1 = \Lambda_2 $, thereby making both terms relevant at the same energy scale.

As an illustrative example, we derive the baryon asymmetry for Eq.~\eqref{eq:-T term}, with the asymmetries for the other interaction terms obtained through analogous steps. In a spatially flat FLRW Universe, the torsion scalar is homogeneous, so that the interaction term in Eq.~\eqref{L_int} reduces to
\begin{equation}
    \mathcal{L}_{\text{original}}=\frac{1}{M_*^2}\partial_\mu(T)J^\mu=\frac{1}{M_*^2}\Dot{T}(n_B-\Bar{n}_B)\label{L_int2}\, .
\end{equation}

Here $J^0=n_B-\Bar{n}_B$ where $n_B$ and $\Bar{n}_B$ are the baryon and anti-baryon number densities respectively. This interaction term \eqref{L_int2} induces an energy shift of baryons over anti-baryons $\sim 2\Dot{T}/M_*^2$. This energy shift results in CPT symmetry violation, and consequently, the chemical potential of baryons is given by
\begin{equation}
    \mu_B=-\Bar{\mu}_B=-\frac{\Dot{T}}{M_*^2}\, .
\end{equation}

Under conditions of thermal equilibrium, in the high-temperature regime where $ \mathbf{T} \gg m $, the net baryon number density is given by~\cite{TheEarlyUniverse}  
\begin{eqnarray}  
    n_b = \frac{g_b}{2\pi^2} \int d^3p \left[ \frac{1}{e^{ \frac{(p - \mu_B)}{\mathbf{T}}} + 1} - \frac{1}{e^{\frac{(p + \mu_B)}{\mathbf{T}}} + 1} \right] \, ,  
\end{eqnarray}  
where $ g_b $ represents the number of internal degrees of freedom of baryons. In a homogeneous and isotropic Universe, this expression simplifies to  
\begin{equation}  
    n_b = \frac{g_b \mathbf{T}^3}{6\pi^2} \left( \pi^2 \frac{\mu_B}{\mathbf{T}} + \left(\frac{\mu_B}{\mathbf{T}}\right)^3 \right) \, .  
\end{equation}  

The conservation of entropy allows the comoving number density of a given particle species to be conveniently expressed as the ratio of the particle number density to the entropy density. Formally, this is defined as  
\begin{equation}  
    N_i \equiv \frac{n_i}{s} \, ,  
\end{equation}  
where the entropy density is given by  
\begin{equation}  
    s= \frac{2\pi^2}{45}g_{\ast s}(\mathbf{T}) \mathbf{T}^3 \, ,  
\end{equation}  
with $ g_{\ast s}(\mathbf{T}) $ defined as  
\begin{equation}  
    g_{\ast s}(\mathbf{T}) = \sum_{i=\text{bosons}} g_i \left(\frac{\mathbf{T}_i}{\mathbf{T}}\right)^3 + \frac{7}{8}\sum_{i=\text{fermions}} g_i \left(\frac{\mathbf{T}_i}{\mathbf{T}}\right)^3 \, .  
\end{equation}  

Assuming that the asymmetry is generated in thermal equilibrium, we can reasonably take $ g_{\ast s} \approx g_{\ast} $, where $ g_{\ast} $ represents the total effective number of relativistic degrees of freedom~\cite{TheEarlyUniverse}. Therefore, if particles are neither produced nor destroyed, as is the case after baryogenesis has concluded, the number density $n_i$ scales with the inverse cube of the scale factor, $n_i \propto a^{-3}$, and $N_i$ constant. This provides a well-defined parameter for observational purposes.  

In the context of an expanding Universe, under the relativistic limit ($ \mathbf{T} \gg m $) and assuming $ \mathbf{T} \gg \mu_B $, the baryon asymmetry produced by the interaction described in Eq. \eqref{eq:-T term} at the temperature at which baryon-number-violating processes freeze out, $ \mathbf{T}_D $, is given by  
\begin{equation} \label{eq:asymmetry T}  
    \frac{n_b}{s} \simeq - \frac{15g_b}{4\pi^2 g_{\ast}} \frac{\dot{T}}{M^2_\ast \mathbf{T}} \Bigg|_{\mathbf{T}_D} \, ,  
\end{equation}  
where $ \mathbf{T} $ denotes the temperature of the state, $ g_{\ast} \approx 106 $ corresponds to the effective number of relativistic degrees of freedom, $ \mathbf{T}_D $ is the decoupling temperature, and $ g_b = \mathcal{O}(1) $ represents the total intrinsic degrees of freedom of baryons.  

The free parameters $M_*$ and $\mathbf{T}_D$ will be chosen as follows: It is assumed that baryogenesis occurs in the radiation-dominated era. This epoch follows inflation whose decoupling temperature is approximately equal to the inflationary scale $M_I$, where $M_I\simeq 1.16\times 10^{16}$ GeV representing the upper bound based on tensor model fluctuation constraints \cite{Planck:2018jri}. As a consequence, if baryogenesis occurs in the radiation era post inflation, the decoupling temperature satisfies the condition $\mathbf{T}_D < \mathbf{T}_{RD} < M_I$, where $\mathbf{T}_{RD}$ is the temperature at which the universe becomes radiation-dominated. The value of $\mathbf{T}_{RD}$ is model dependent and as such, in this work we will not assume any inflationary model but consider the estimate $\mathbf{T}_{RD}\lesssim10^9-10^{14}$ GeV \cite{Amin:2014eta,Bezrukov:2007ep,Bezrukov:2011gp}. For the lower bound, we will assume that baryogenesis occurs in the early stages of the radiation epoch with $\mathbf{T}_D \gtrsim 10^7$ GeV. Hence, the decoupling temperature is in the interval $10^7 \lesssim \mathbf{T}_D \lesssim 10^{14}$ GeV. However, in order to avoid the production of gravitons, temperatures $\mathbf{T}_D < 10^{10}$ GeV will be favored \cite{Kohri:2005wn}.

Regarding the parameter $M_*$, in the original work~\cite{Davoudiasl:2004gf} it is noted that $M_*$ does not necessarily need to reach the Planck scale $M_{\rm Pl}$. If the baryon-number violation is soft, it can be characterized by the Majorana mass $M_R$ of the right-handed neutrino. The interaction in Eq.~\eqref{L_int} does not lead to any violation of unitarity up to the Planck scale, even when $M_*^2 = M_R M_{\rm Pl}$. Consequently, the cutoff scale can be chosen lower in order to reproduce the observed baryon-to-entropy ratio.

It is natural to replace the Ricci scalar $R$ in the interaction term \eqref{L_int} with the negative torsion scalar $-T$. This type of coupling has been employed in studies of baryogenesis within $f(T)$ gravity models~\cite{f(T)baryogenesis}. More sophisticated extensions, such as $f(T,\mathcal{T})$, where $\mathcal{T}$ denotes the trace of the energy–momentum tensor, require the inclusion of this additional quantity in the interaction term, as illustrated in~\cite{Mishra:2023khd}. In the context of our $f(T,L_m)$ model, the inclusion of the term $\partial_\mu(-L_m)$ in the interaction is essential to achieve the observed baryon-to-entropy ratio while maintaining physically viable values for the free parameters of each model, as will be demonstrated in the following subsections.

\subsubsection{Assumptions and limitations}

In the following analysis we make several standard assumptions, which we briefly summarize here. We work in a spatially flat FLRW background with a power-law scale factor 
\begin{equation}\label{scale factor}
a(t) = A t^n\,,
\end{equation}
taking $n = 1/2$ and an equation-of-state parameter $w = 1/3$ to model the radiation-dominated epoch. The matter Lagrangian is chosen as $L_m = P = w \rho$, which is the usual choice in modified-gravity models with explicit matter--geometry couplings~\cite{fRLm,Harko:2014gwa,Harko:2013yb,Harko:2014sja}. We further assume that the Universe passes through a sequence of quasi-equilibrium states, so that the radiation energy density obeys the standard relation
\begin{equation}\label{Decoupling Temperature}
\rho = (\pi^2 g_\ast / 30)\,\mathbf{T}^4,
\end{equation}
and we parameterize baryon-number freeze-out by a decoupling temperature $\mathbf{T}_D$ in the range $10^7 \lesssim \mathbf{T}_D \lesssim 10^{14}\,\mathrm{GeV}$, consistent with typical reheating scenarios~\cite{Planck:2018jri,Amin:2014eta,Bezrukov:2007ep,Bezrukov:2011gp,Kohri:2005wn}. Finally, we impose the constraint $|\rho_{\rm MG}/\rho_r| \lesssim 0.10$, so that the modified-gravity contribution remains a small correction to standard radiation during baryogenesis. Relaxing these assumptions, for instance by allowing for non-standard equations of state or more general fluids, would certainly be interesting, but lies beyond the scope of the present work.

We will study three different types of couplings between the baryonic current and the gravitational sector: the coupling to torsion alone (Sec.~\ref{Coupling_T}); the coupling to torsion together with the matter Lagrangian (Sec.~\ref{Coupling_T_Lm}); and finally, the coupling to a general function $f(T,L_m)$ (Sec.~\ref{Coupling_f}). These interaction terms allow us to investigate the effects of non-minimal couplings and their implications for the generation of baryon asymmetry, providing insight into the complex interplay between fundamental scalar fields and baryonic processes in the early Universe. Moreover, they offer new perspectives for exploring the connections between gravitational dynamics and particle physics.

\subsection{Coupling between $\partial_\mu (-T)$ and $J^\mu$}\label{Coupling_T}

\subsubsection{$f=BTL_m$}\label{BTLm1}

Let us start by considering a model with a non-minimal coupling between the matter sector and the torsion scalar $f=BTL_m$, where $B$ is a free parameter. 
Physically, $B$, with units of $\text{mass}^{-4}$, quantifies the strength of the direct torsion–matter
coupling: the limit $B \to 0$ recovers standard teleparallel gravity, while non-zero values
of $B$ parameterize small deviations from GR. Substituting $f=BTL_m$ into Eq.~(\ref{first Friedmann}) and using Eq.~(\ref{T}), we obtain the energy density
\begin{equation}
    \rho=\left[\frac{t^2}{3n^2M_{\text{Pl}}^2}-B(2w+1)\right]^{-1}\,.
    \label{BTL_m_rho}
\end{equation}

We can relate the time of decoupling $t_D$ with the temperature of decoupling by using Eq. (\ref{Decoupling Temperature}) that relates the
total radiation density with the energy of all relativistic species giving
\begin{equation}
    t_D=\left[3Bn^2M_{\text{Pl}}^2(2w+1)+\frac{90n^2M_{\text{Pl}}^2}{\pi^2g_*\mathbf{T}_D^4}\right]^{1/2}\,.
   \label{BTL_m_tD} 
\end{equation}

Using this previous result, the torsion scalar (\ref{T}) and taking into account Eq. \eqref{eq:asymmetry T}, the resulting baryon-to-entropy ratio becomes
\begin{equation}
    \frac{n_b}{s}\simeq\frac{180n^2g_b\epsilon}{4\pi^2 g_* M^2_* \mathbf{T}_D}\left[3Bn^2M_{\text{Pl}}^2(2w+1)+\frac{90n^2M_{\text{Pl}}^2}{\pi^2g_*\mathbf{T}_D^4}\right]^{-3/2}\, .
\end{equation}

\begin{figure}[t!]
    \centering
    \includegraphics[width=0.475\textwidth]{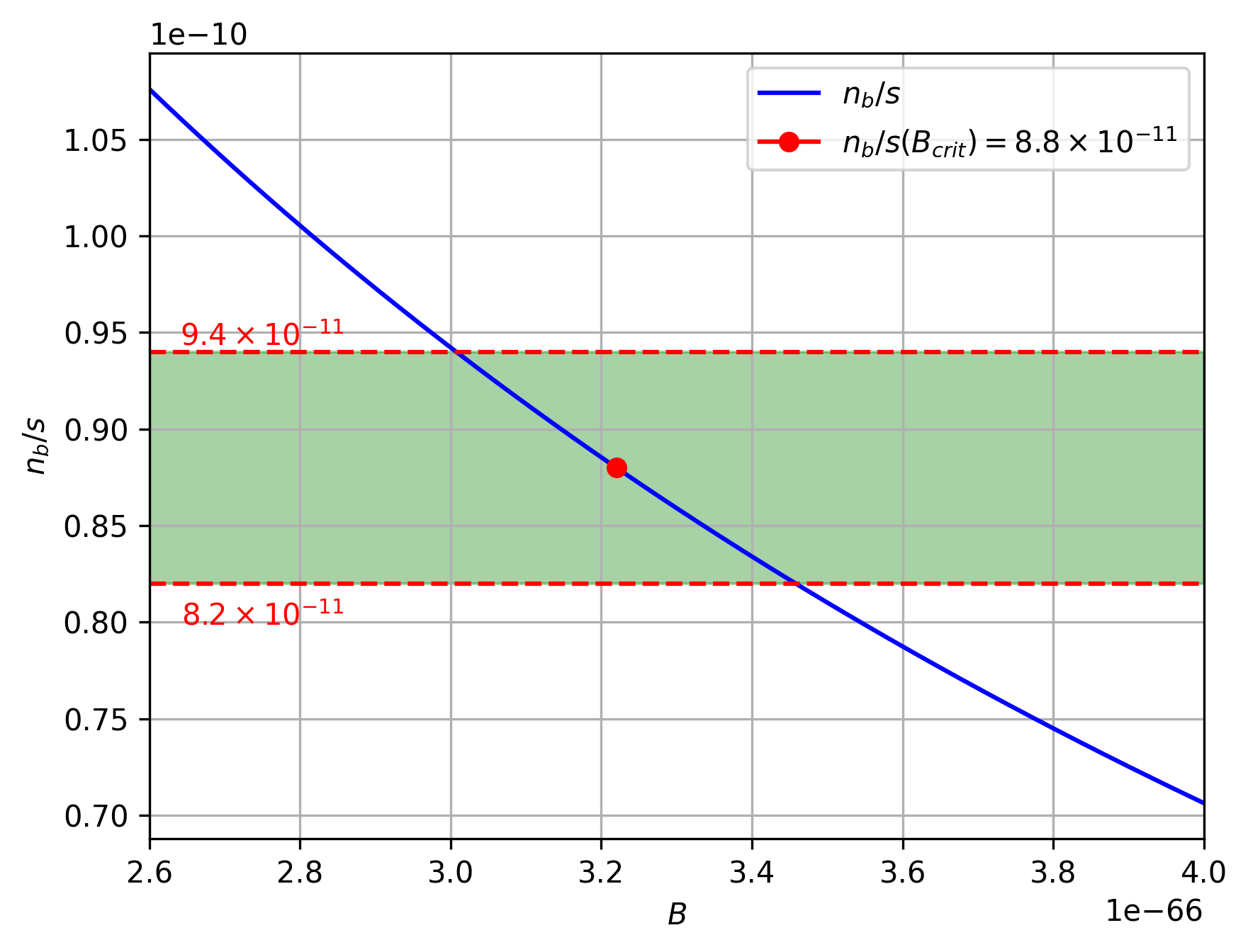}
    \caption{Plot of the baryon to entropy ratio for model $f=BTL_m$ for varying B, $\mathbf{T}_D=10^{16}\ GeV$, $M_*=2.4\times 10^{17}\ \text{GeV}$, $\epsilon=1$, $n=\frac{1}{2}$ and $w=\frac{1}{3}$.}
    \label{fig: BTLm1}
\end{figure}

Figure~\ref{fig: BTLm1} illustrates the dependence of the baryon-to-entropy ratio on the parameter $B$. The blue curve shows the evolution of $\frac{n_b}{s}(B)$ as $B$ varies, while the dashed red lines denote the upper and lower bounds imposed by current observational constraints. The red dot indicates the critical value of $B$, with $B_{\rm crit} \approx 3.22 \times 10^{-66}\,\mathrm{GeV}^{-4}$. The true coupling constant for this model is $BM_{\text{Pl}}^2\approx1.85\times10^{-29}$. The new term arising from the contribution of modified gravity is much smaller then the torsion scalar $T$. This value implies that a small contribution from the additional $BTL_m$ term is sufficient to obtain an asymmetry in accordance with observational data. Additionally, for this particular model, achieving successful baryogenesis requires a decoupling temperature of the order $\mathbf{T}_D \sim M_I$, which lies beyond the estimated temperature range considered in this study being similar to what occurs in the original gravitational baryogenesis interaction term~\cite{Davoudiasl:2004gf}. Consequently, temperatures within the considered range are insufficient to generate the observed baryon-to-entropy ratio, highlighting the sensitivity of baryogenesis to both the coupling parameter $B$ and the decoupling temperature.

\subsubsection{$f=B(1-e^{-p\sqrt{|T|}})L_m$}\label{sqrt1}

This model arises from a coupling between the well-studied $f(T)$ ``Linder model''~\cite{Linder} and the matter Lagrangian. The corresponding baryogenesis scenario for this $f(T)$ model has been previously analyzed in~\cite{f(T)baryogenesis}. The model is characterized by two parameters, $B$ and $p$. Here $B$ again controls the overall strength of the torsion–matter coupling and has mass
dimension $[B]=\mathrm{mass}^{-2}$, so that $f$ has the correct dimension. The new parameter
$p$ enters in the dimensionless combination $p\sqrt{|T|}$ and therefore has dimension
$[p]=\mathrm{mass}^{-1}$. It introduces a characteristic torsional (or energy) scale
$|p|^{-1}$: for $|p\sqrt{|T|}| \ll 1$ the exponential can be expanded and the model reduces
to a linear perturbation around GR, while for $|p\sqrt{|T|}|\gtrsim 1$ the non-linear
torsion–matter effects become significant. By substituting $f = B \left(1 - e^{-p \sqrt{|T|}}\right) L_m$
into Eq.~(\ref{first Friedmann}) and employing Eq.~(\ref{T}), we obtain the expression for the energy density
\begin{equation}
    \rho\approx\left[\frac{t^2}{3n^2M_{\text{Pl}}^2}+\frac{Bp}{\sqrt{6}n}(1-w)t+Bp^2w\right]^{-1}\, ,
    \label{sqrt_rho}
\end{equation}
where we considered the approximation $e^{-p\sqrt{|T|}}\approx1-p\sqrt{|T|}$. Using Eq. \eqref{Decoupling Temperature} we find the decoupling time $t_D$ to be given by
\begin{eqnarray}
	&&t_D = \left({\frac{2}{3 n^2 M_{\rm Pl}^2}}\right)^{-1}
		\Bigg[-\frac{B p}{\sqrt{6}\, n}(1-w)
			\nonumber \\
	&& \hspace{-1.0cm}	+ \sqrt{\frac{B^2 p^2}{6 n^2} (1-w)^2 
			- \frac{4}{3 n^2 M_{\rm Pl}^2} 
			\left( B p^2 w - \frac{30}{\pi^2 g_* \mathbf{T}_D^4} \right)}\, \Bigg]
	 ,
	\label{sqrt_tD}
\end{eqnarray}
and using the torsion scalar (\ref{T}) and Eq. \eqref{eq:asymmetry T} the resulting baryon-to-entropy ratio becomes
\begin{equation}
    \begin{split}
        &\frac{n_b}{s}\simeq\frac{10\sqrt{6}g_b\epsilon}{\pi^2 g_*nM_{\text{Pl}}^3 M^2_* \mathbf{T}_D}\left[-\frac{Bp}{\sqrt{6}n}(1-w)+\right.\\
        &\left.\sqrt{\frac{B^2p^2}{6n^2}(1-w)^2-\frac{4}{3n^2M_{\text{Pl}}^2}\left(Bp^2w-\frac{30}{\pi^2 g_* \mathbf{T}_D^4}\right)}\right]^{-3}\, .
    \end{split}
\end{equation}

\begin{figure}[ht]
	\centering
	\includegraphics[width=0.475\textwidth]{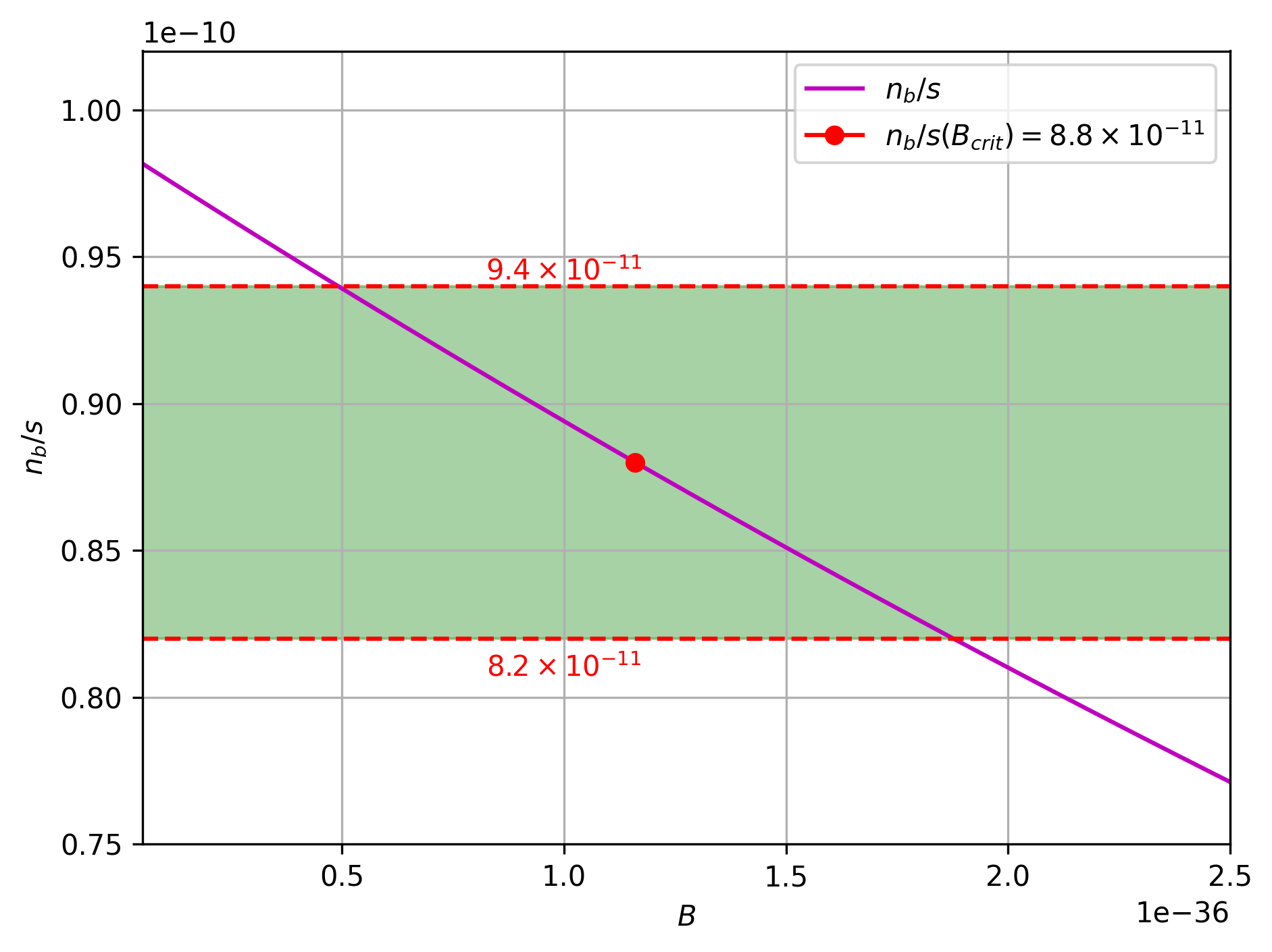}
	\caption{Plot of the baryon to entropy ratio for model $f=B(1-e^{-p\sqrt{|T|}})L_m$ for varying B, $\mathbf{T}_D=10^{16}\ \text{GeV}$, $M_*=5\times 10^{17}\ \text{GeV}$, $p=-10^{-16}\, \mathrm{GeV}^{-1}$ , $\epsilon=1$, $n=\frac{1}{2}$ and $w=\frac{1}{3}$.}
	\label{fig:sqrt1}
\end{figure}

Figure~\ref{fig:sqrt1} shows the dependence of the baryon-to-entropy ratio on the parameter $B$. The pink curve represents the evolution of $\frac{n_b}{s}(B)$, while the dashed red lines indicate the upper and lower bounds imposed by current observational constraints. The red dot denotes the critical value of $B$, given by $B_{\rm crit} \approx 1.160 \times 10^{-36}\,\mathrm{GeV}^{-2}$. For this model, the effective coupling constant is $BM_{\text{Pl}}^2 \approx 6.68$. 
Successful baryogenesis within this framework requires a decoupling temperature of order $\mathbf{T}_D \sim M_I$, which lies beyond the temperature range considered in this study. Consequently, the explored range of temperatures is insufficient to reproduce the observed baryon-to-entropy ratio.

\subsubsection{$f=B(1-e^{-pT})L_m$}\label{exp1}

This model arises from a coupling between a previously studied $f(T)$ model~\cite{exponential} and the matter Lagrangian. The baryogenesis scenario for this $f(T)$ model has also been investigated in~\cite{f(T)baryogenesis}. The model is characterized by the parameters $B$ and $p$. The $p$ parameter has now dimension $[p]=\mathrm{mass}^{-2}$ and the scale $|p|^{-1/2}$ indicates the torsional/energy
scale at which the exponential deformation departs from the GR limit. For $|pT|\ll1$ the
model effectively reduces to a linear correction $f\simeq BpTL_m$, whereas for $|pT|\gtrsim1$
the exponential structure leads to stronger modifications of the early-Universe dynamics. By substituting $f = B \left( 1 - e^{-p T} \right) L_m$
into Eq.~(\ref{first Friedmann}) and using Eq.~(\ref{T}), we obtain the corresponding expression for the energy density
\begin{equation}
    \rho\approx\left[\frac{t^2}{3n^2M_{\text{Pl}}^2}-Bp(2w+1)-12Bwp^2n^2t^{-2}\right]^{-1}\, .
    \label{exp_rho}
\end{equation}

Using Eq. (\ref{Decoupling Temperature}) we find the decoupling time $t_D$
\begin{equation}
    \begin{split}
        &t_D=\left[\frac{2}{3n^2M_{\text{Pl}}^2}\right]^{-1/2}\left(Bp(2w+1)+\frac{30}{\pi^2 g_*\mathbf{T}_D^4}\right.\\
        &\left. +\sqrt{\left(Bp(2w+1)+\frac{30}{\pi^2 g_*\mathbf{T}_D^4}\right)^2+\frac{16Bwp^2}{M_{\text{Pl}}^2}}\right)^{1/2}
    \end{split}\, ,\label{exp_tD}
\end{equation}
and taking into account the torsion scalar (\ref{T}) and Eq. \eqref{eq:asymmetry T}, the resulting baryon-to-entropy ratio becomes
\begin{equation}
    \begin{split}
        &\frac{n_b}{s}\simeq\frac{10\sqrt{6}g_b\epsilon}{\pi^2 g_*nM_{\text{Pl}}^3 M^2_* \mathbf{T}_D}\left(Bp(2w+1)+\frac{30}{\pi^2 g_*\mathbf{T}_D^4}\right.\\
        &\left. +\sqrt{\left(Bp(2w+1)+\frac{30}{\pi^2 g_*\mathbf{T}_D^4}\right)^2+\frac{16Bwp^2}{M_{\text{Pl}}^2}}\right)^{-3/2}
    \end{split}
\end{equation}

\begin{figure}[t!]
	\centering
	\includegraphics[width=0.475\textwidth]{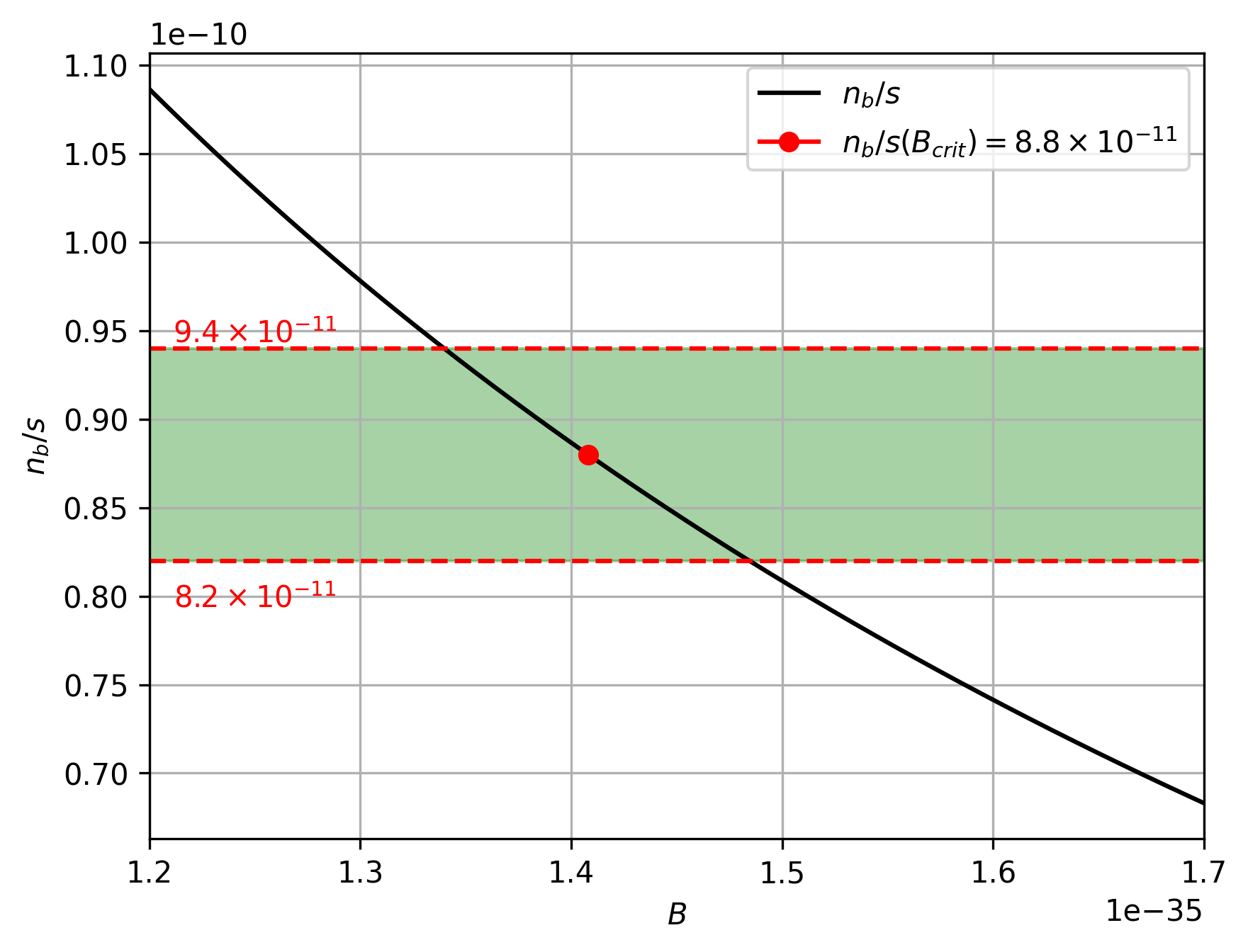}
	\caption{Plot of the baryon to entropy ratio for model $f=B(1-e^{-pT})L_m$ for varying B, $\mathbf{T}_D=10^{16}\ \text{GeV}$, $M_*=1\times 10^{17}\ GeV$, $p=10^{-30}\, \mathrm{GeV}^{-2}$, $\epsilon=1$ $n=\frac{1}{2}$ and $w=\frac{1}{3}$.}
	\label{fig:exp1}
\end{figure}

Figure~\ref{fig:exp1} illustrates how the baryon-to-entropy ratio varies with the parameter $B$. The black curve traces the behaviour of $\frac{n_b}{s}(B)$, while the dashed red lines represent the observationally allowed upper and lower bounds. The red dot identifies the critical value of $B$, namely $B_{\rm crit} \approx 1.408 \times 10^{-35}\,\mathrm{GeV}^{-2}$. For this model, the effective coupling constant takes the value $BM_{\text{Pl}}^2 \approx 81.1$. 
In this framework, successful baryogenesis requires a decoupling temperature on the order of $\mathbf{T}_D \sim M_I$, which lies outside the temperature interval considered in this analysis. Hence, within the examined range, the predicted baryon-to-entropy ratio falls short of the observed value.

\subsection{Coupling between $\partial_\mu(-T-\frac{L_m}{L_0})$ and $J^\mu$} \label{Coupling_T_Lm}

For the CPT-violating term~(\ref{CP-Violating}), the induced chemical potential takes the form  $\mu \sim \pm \frac{\Dot{T} + \partial_t(L_m)/L_0}{M_*^2}$, where $M_*$ denotes the cutoff scale of the effective interaction. Consequently, the corresponding baryon-to-entropy ratio can be expressed as
\begin{equation}
    \frac{n_b}{s}\simeq \frac{15g_b\epsilon}{4\pi^2g_*}\frac{\Dot{T}+\frac{\partial_t(L_m)}{L_0}}{M^2_*\mathbf{T}}\Bigg{|}_{\mathbf{T}_D}\,.
\end{equation}

\subsubsection{$f=BTL_m$\label{BTL_m}}

Using the energy density \eqref{BTL_m_rho} and the decoupling time \eqref{BTL_m_tD} found in \eqref{BTLm1}, we can determine the baryon-to-entropy ratio associated with the CPT-violating interaction term in Eq.~\eqref{CP-Violating}
\begin{eqnarray}
        \frac{n_b}{s}& \approx &\frac{15g_b\epsilon}{4\pi^2 g_* M^2_* \mathbf{T}_D}\bigg[12n^2t_D^{-3}
        	\nonumber \\
        &&-\left(\frac{\pi^2g_*\mathbf{T}_D^4}{30}\right)^{2}\frac{2w}{3M^2_{pl}M_*^2n^2}t_D\Bigg]\,.
    \label{BTL_m_baryogenesis}
\end{eqnarray}

\begin{figure}[t!]
    \centering
    \includegraphics[width=0.475\textwidth]{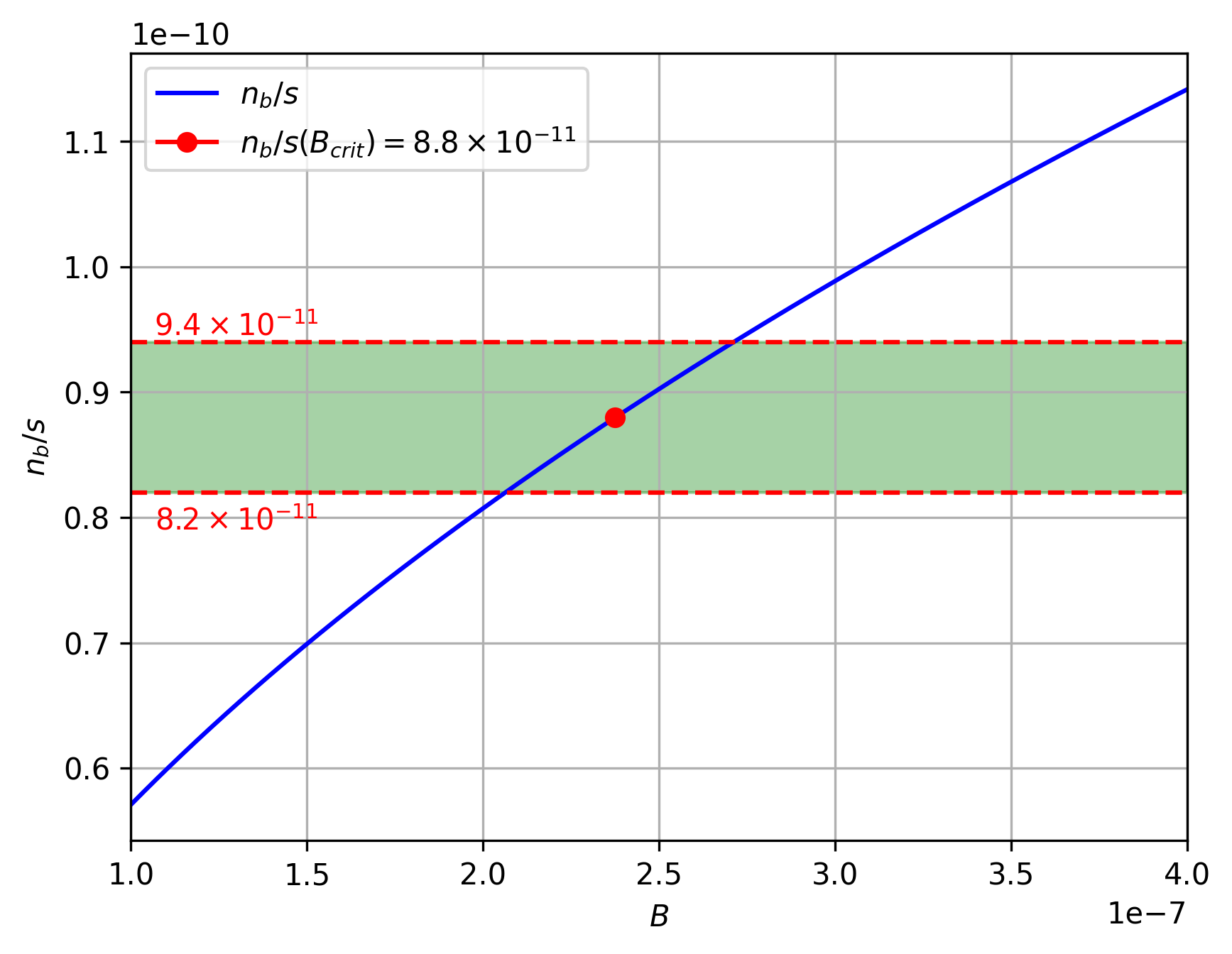}
    \caption{Plot of the baryon to entropy ratio for model $f=BTL_m$ for varying B, $\mathbf{T}_D=10^{9}\ \text{GeV}$, $M_*=1\times 10^{13}\ \text{GeV}$, $\epsilon=-1$, $n=\frac{1}{2}$ and $w=\frac{1}{3}$.}
    \label{fig:BTL_m}
\end{figure}

Figure~\ref{fig:BTL_m} shows the dependence of the baryon-to-entropy ratio on the parameter $B$. The blue curve represents the evolution of $\frac{n_b}{s}(B)$, the dashed red lines indicate the upper and lower bounds imposed by observational constraints, and the red dot marks the critical value of $B$, with $B_{\rm crit} \approx 2.375 \times 10^{-7}\,\mathrm{GeV}^{-4}$. 

\subsubsection{$f=B(1-e^{-p\sqrt{|T|}})L_m$\label{sqrt}}

Using the energy density \eqref{sqrt_rho} and the decoupling time \eqref{sqrt_tD} found in Sec. \ref{sqrt1}, we find the baryon-to-entropy ratio for the CPT-violating interaction term \eqref{CP-Violating}
\begin{eqnarray}
        \frac{n_b}{s}\simeq\frac{15g_b\epsilon}{4\pi^2g_*M_*^2\mathbf{T}_D}
        \left[12n^2t_D^{-3}-\frac{w}{M_*^2}\left(\frac{\pi^2g_*\mathbf{T}_D^4}{30}\right)^2\right.
        	\nonumber \\
        \left.\times\left(\frac{2}{3n^2M_{\text{Pl}}^2}t_D+\frac{Bp}{\sqrt{6}n}(1-w)\right)\right]
        \, .
    \label{sqrt1_normal_baryogenesis}
\end{eqnarray}

\begin{figure}[ht]
    \centering
    \includegraphics[width=0.475\textwidth]{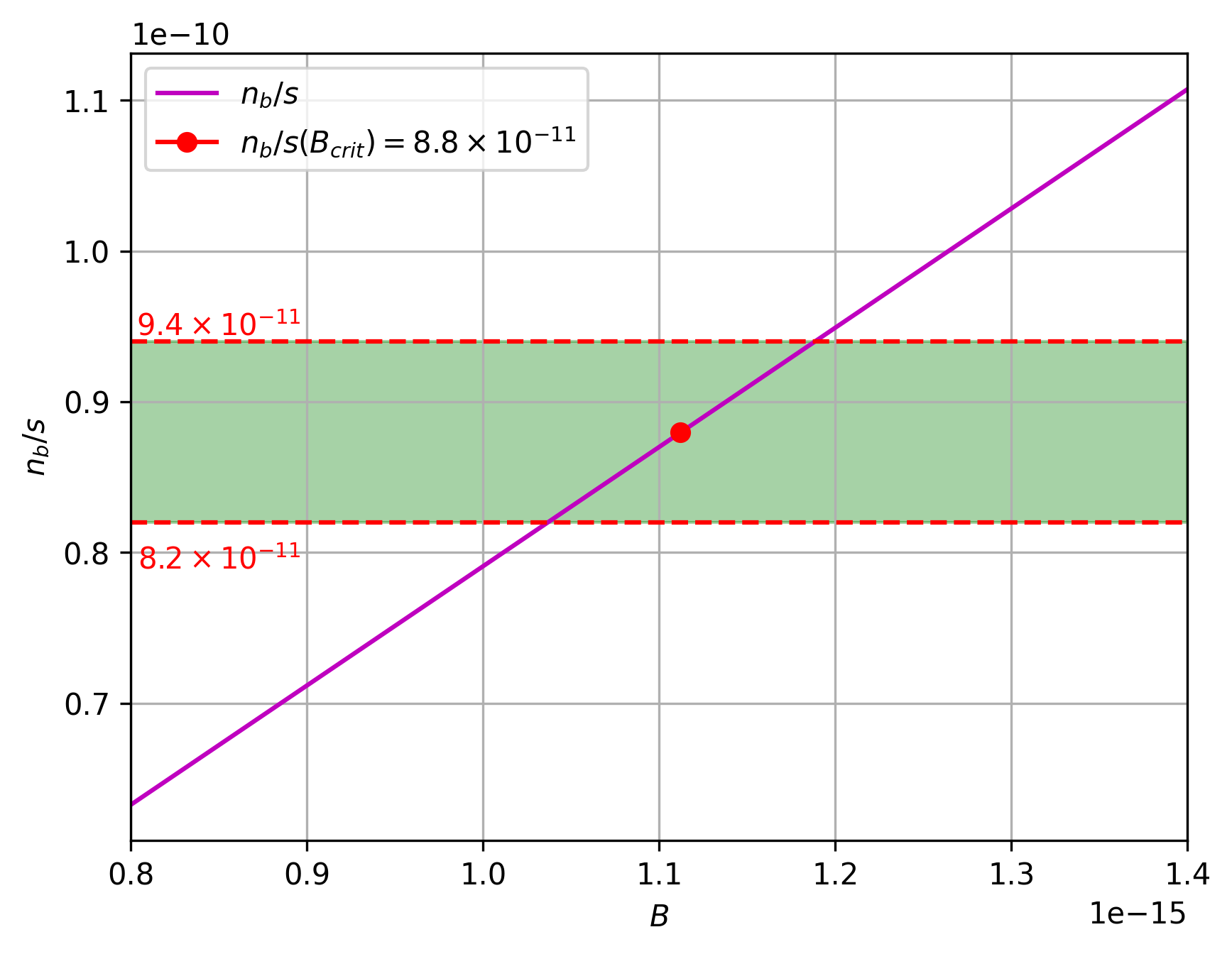}
    \caption{Plot of the baryon to entropy ratio for model $f=B(1-e^{-p\sqrt{|T|}}) L_m$ for varying B,  $\mathbf{T}_D=10^{8}\ \text{GeV}$, $M_*=1\times 10^{10}\ \text{GeV}$, $p=-10^{-11}\ \text{GeV}^{-1}$, $\epsilon=-1$ , $n=\frac{1}{2}$ and $w=\frac{1}{3}$}
    \label{fig:sqrt1_normal}
\end{figure}

Figure~\ref{fig:sqrt1_normal} illustrates the dependence of the baryon-to-entropy ratio on the parameter $B$. The pink curve represents the evolution of $\frac{n_b}{s}(B)$, the dashed red lines indicate the upper and lower bounds imposed by observational constraints, and the red dot marks the critical value of $B$, with $B_{\rm crit} \approx 1.112 \times 10^{-15}\,\mathrm{GeV}^{-2}$. For a fixed value of $M_*$, higher decoupling temperatures correspond to lower values of $B$ and $p$, and conversely, lower temperatures require higher values of these parameters.

\subsubsection{$f=B(1-e^{-pT})L_m$\label{exp}}

Using the energy density \eqref{exp_rho} and the decoupling time \eqref{exp_tD} found in Sec. \ref{exp1} we find the baryon-to-entropy ratio for the CPT-violating interaction term \eqref{CP-Violating}
\begin{eqnarray}
        \frac{n_b}{s}\simeq \frac{15g_b\epsilon}{4\pi^2g_*M_*^2\mathbf{T}_D}\left(12n^2t_D^{-3}-\frac{w}{M_*^2}\left(\frac{\pi^2g_*\mathbf{T}_D^4}{30}\right)^2\right.
        	\nonumber \\
        \left.\times\left[\frac{2}{3n^2M_{\text{Pl}}^2}t_D+24Bwp^2n^2t_D^{-3}\right)\right]\,.
\end{eqnarray}

\begin{figure}[t!]
    \centering
    \includegraphics[width=0.475\textwidth]{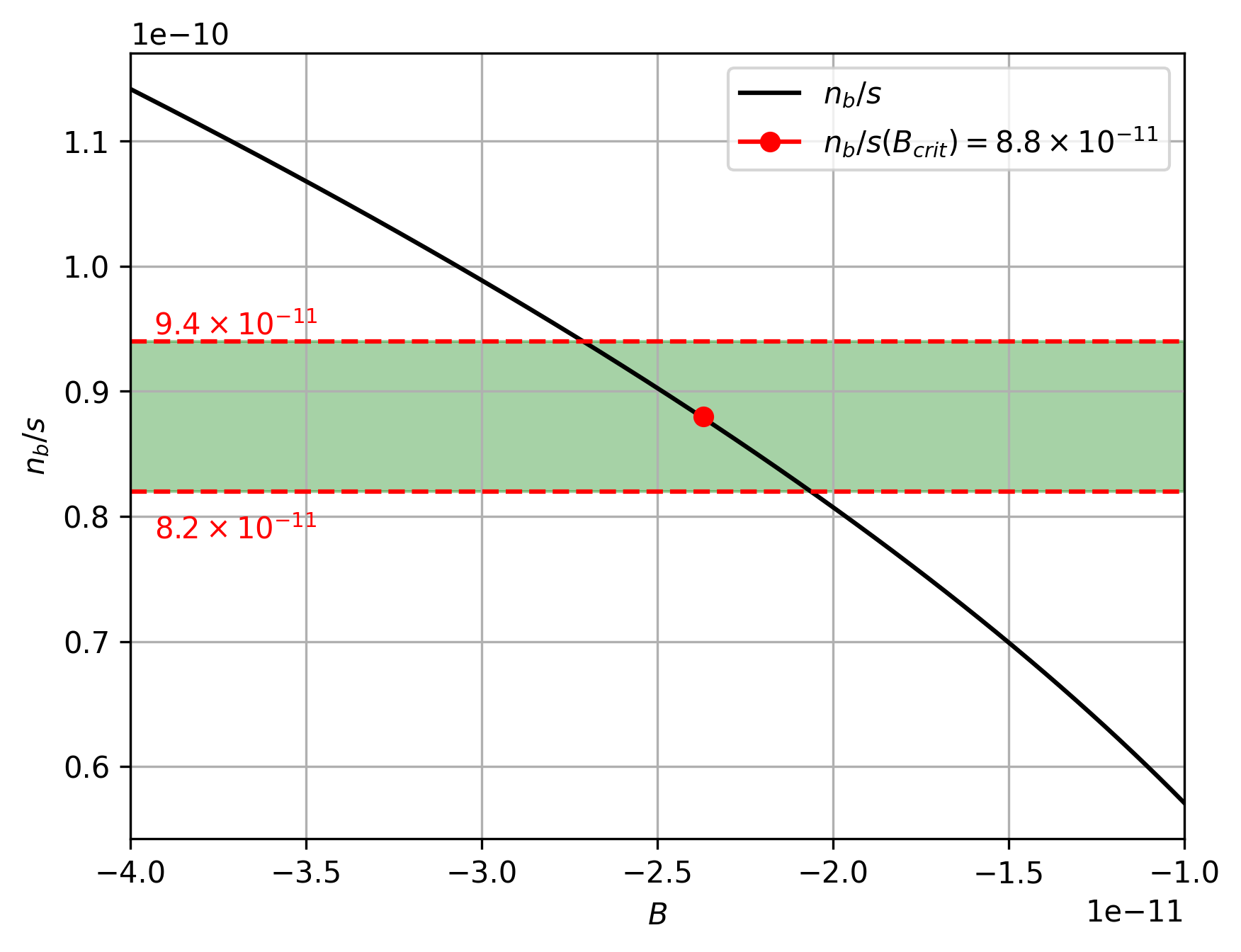}
    \caption{Plot of the baryon to entropy ratio for model $f=B(1-e^{-pT}) L_m$ for varying B,  $\mathbf{T}_D=10^{9}\ \text{GeV}$, $M_*=1\times 10^{11}\ GeV$,  $p=-10^{-12}\ \text{GeV}^{-2}$, $\epsilon=-1$ , $n=\frac{1}{2}$ and $w=\frac{1}{3}$}
    \label{fig:exp_1_normal}
\end{figure}

Figure~\ref{fig:exp_1_normal} illustrates the dependence of the baryon-to-entropy ratio on the parameter $B$. The black curve shows the evolution of $\frac{n_b}{s}(B)$, while the dashed red lines denote the observationally allowed upper and lower bounds. The red dot corresponds to the critical value of $B$, given by $B_{\rm crit} \approx -2.370 \times 10^{-11}\,\mathrm{GeV}^{-2}$. For a fixed value of $M_*$, higher decoupling temperatures are associated with smaller absolute values of both $B$ and $p$, whereas lower decoupling temperatures require larger absolute values of these parameters.

\subsubsection{Discussion\label{Discussion1}}

For the models considered here, and in contrast to the results discussed in Sec.~\ref{Coupling_T}, the inclusion of the $\partial_\mu(-L_m)$ interaction term provides greater flexibility in the choice of free parameters. Specifically, it allows the baryon-to-entropy ratio to reach the observed value for a wider range of $M_*$, $\mathbf{T}_D$, $B$, and $p$. Analyzing now the parameters of the asymmetry, for the $f = B T L_m$ model~\eqref{BTL_m}, taking $\mathbf{T}_D = 10^7\,\mathrm{GeV}$ and $M_* = 10^{10}\,\mathrm{GeV}$ yields $B_{\rm crit} \approx 2.375 \times 10^{-3}\,\mathrm{GeV}^{-4}$. Maintaining this same order-of-magnitude difference between $\mathbf{T}_D$ and $M_*$, consecutive decoupling temperatures result in $B_{\rm crit}$ values differing by roughly six orders of magnitude. Since higher temperatures correspond to smaller values of $B$, this indicates that $B$ decreases as $\mathbf{T}_D$ increases.

Similarly, for the $f = B \left(1 - e^{-p \sqrt{|T|}}\right) L_m$ model \eqref{sqrt}, taking $\mathbf{T}_D = 10^7\,\mathrm{GeV}$, $M_* = 10^{10}\,\mathrm{GeV}$, and $p = -10^{-11}\,\mathrm{GeV}^{-1}$, we obtain $B_{\rm crit} \approx 1.112 \times 10^{-8}\,\mathrm{GeV}^{-2}$. Maintaining the same order-of-magnitude difference between $\mathbf{T}_D$ and $M_*$ and $p\sqrt{|T|}$ fixed for all temperatures, consecutive decoupling temperatures yield $B_{\rm crit}$ values differing by approximately three orders of magnitude, with $B$ decreasing for higher temperatures.  
For the model $f = B \left(1 - e^{-p T}\right) L_m$~\eqref{exp}, with $\mathbf{T}_D = 10^7\,\mathrm{GeV}$, $M_* = 10^9\,\mathrm{GeV}$, and $p = -10^{-12}\,\mathrm{GeV}^{-2}$, we find $B_{\rm crit} \approx -2.370 \times 10^{1}\,\mathrm{GeV}^{-2}$. Applying the same reasoning, but with the difference between $\mathbf{T}_D$ and $M_*$ now being two orders of magnitude and $pT$ valued fixed for all temperatures, results in $B_{\rm crit}$ values that are approximately six orders of magnitude smaller for consecutive temperatures. The value of $p$ changes with temperature to ensure the validity of the approximations considered in the derivation of ~\eqref{sqrt_rho} and ~\eqref{exp_rho}.

For all three models, with the $B$ values found, the true coupling $BM_{\text{Pl}}^2$ implies the new modified gravity terms are highly impactful, specially for lower temperatures. These results entail $\rho_r \simeq -\rho_{\rm MG}$, yielding a very small, yet positive, Hubble parameter. If one wishes to obtain higher values for the Hubble parameter, or to have $\rho_{\rm MG}$ constitute only a small modification to GR, smaller values of $B$ than those presented above are required. Achieving this, while simultaneously supporting successful baryogenesis, requires considering $M_*$ approximately one order of magnitude larger than $\mathbf{T}_D$. Let us now examine the scenario in which $\rho_{\rm MG}$ represents a small correction to GR. The corresponding constraints for the models \eqref{BTL_m}, \eqref{sqrt}, and \eqref{exp} are, respectively,
\begin{equation}
    \left|\frac{\rho_{MG}}{\rho_r}\right|=3M_{\text{Pl}}^2B(2w+1)H^2\, ,
\end{equation}
\begin{equation}
    \left|\frac{\rho_{MG}}{\rho_r}\right|=3M_{\text{Pl}}^2\left(\frac{Bp}{\sqrt{6}}(1-w)H+Bp^2wH^2\right)\, ,
\end{equation}
\begin{equation}
    \left|\frac{\rho_{MG}}{\rho_r}\right|=3M_{\text{Pl}}^2\left(Bp(2w+1)H^2+12Bwp^2H^4\right)\, .
\end{equation}

For all three models, we found that only decoupling temperatures $\mathbf{T}_D \geq 10^{12}\,\mathrm{GeV}$ simultaneously support successful baryogenesis and satisfy $\left|\frac{\rho_{\rm MG}}{\rho_r}\right| \lesssim 0.10$.  
For the $BTL_m$ model, at $\mathbf{T}_D = 10^{12}\,\mathrm{GeV}$ and $B = 2.204 \times 10^{-51}\,\mathrm{GeV}^{-4}$, we obtain $n_b/s \approx 2.51 \times 10^{-11}$, with the baryon-to-entropy ratio converging to $2.37 \times 10^{-11}$ for smaller values of $B$. At $\mathbf{T}_D = 10^{13}\,\mathrm{GeV}$, similar behaviour is observed: $n_b/s \approx 2.51 \times 10^{-10}$ for $B = 2.204 \times 10^{-55}\,\mathrm{GeV}^{-4}$, converging to $2.37 \times 10^{-10}$ for lower $B$. Finally, with $M_* = 2 \times 10^{15}\,\mathrm{GeV}$ and $\mathbf{T}_D = 10^{14}\,\mathrm{GeV}$, we find $n_b/s \approx 1.57 \times 10^{-10}$ for $B = 2.204 \times 10^{-59}\,\mathrm{GeV}^{-4}$, converging to $1.48 \times 10^{-10}$ for smaller values of $B$.

For the model $f = B \left(1 - e^{-p \sqrt{|T|}}\right) L_m$, at $\mathbf{T}_D = 10^{12}\,\mathrm{GeV}$ and $p = -10^{-11}\,\mathrm{GeV}^{-1}$, we find $n_b/s \approx 2.37 \times 10^{-11}$ for $B = 1.112 \times 10^{-33}\,\mathrm{GeV}^{-2}$. At $\mathbf{T}_D = 10^{13}\,\mathrm{GeV}$ with the same $p$, we obtain $n_b/s \approx 2.37 \times 10^{-10}$ for $B = 1.112 \times 10^{-35}\,\mathrm{GeV}^{-2}$. Finally, for $\mathbf{T}_D = 10^{14}\,\mathrm{GeV}$, $M_* = 2 \times 10^{14}\,\mathrm{GeV}$, and $p = -10^{-12}\,\mathrm{GeV}^{-1}$, we find $n_b/s \approx 1.48 \times 10^{-10}$ for $B = 1.112 \times 10^{-37}\,\mathrm{GeV}^{-2}$. Smaller values of $B$ than those indicated here yield the same baryon-to-entropy ratio.

For the model $f = B \left(1 - e^{-p T}\right) L_m$, we find that for $B = -2.370 \times 10^{-37}\,\mathrm{GeV}^{-2}$, with $\mathbf{T}_D = 10^{12}\,\mathrm{GeV}$ and $p = -10^{-14}\,\mathrm{GeV}^{-2}$, the resulting baryon-to-entropy ratio is $n_b/s \approx 2.51 \times 10^{-11}$. For $\mathbf{T}_D = 10^{13}\,\mathrm{GeV}$ , $B= -2.370\times 10^{-48}\,\mathrm{GeV}^{-2}$ and $p = -10^{-18}\,\mathrm{GeV}^{-2}$, we obtain $n_b/s \approx 2.59 \times 10^{-10}$. Finally, for $\mathbf{T}_D = 10^{14}\,\mathrm{GeV}$, $M_* = 2 \times 10^{15}\,\mathrm{GeV}$, $B= -2,370\times10^{-56}\,\mathrm{GeV}^{-2}$ and $p = -10^{-22}\,\mathrm{GeV}^{-2}$, we find $n_b/s \approx 1.62 \times 10^{-10}$. Smaller values of $B$ lead to convergence at $n_b/s \approx 2.37 \times 10^{-11}$ for $\mathbf{T}_D = 10^{12}\,\mathrm{GeV}$, $n_b/s \approx 2.37 \times 10^{-10}$ for $\mathbf{T}_D = 10^{13}\,\mathrm{GeV}$, and $n_b/s \approx 1.48 \times 10^{-10}$ for $\mathbf{T}_D = 10^{14}\,\mathrm{GeV}$.

The parameter values of $p$ and $B$ adopted in this section satisfy the validity of the approximations 
$e^{-pT} \approx 1 - pT$ and $e^{-p\sqrt{|T|}} \approx 1 - p\sqrt{|T|}$ used in the derivations 
of Eqs.~\eqref{sqrt_rho} and~\eqref{exp_rho}.

These values of $B$ yield baryon-to-entropy ratios below the observational bounds; however, this does not present a problem, as one can adjust $M_*$ to bring $n_b/s$ within the observationally favored range. Interestingly, the three models considered here produce similar baryon-to-entropy ratios with the parameter values obtained. A possible explanation for this similarity lies in the nature of the non-minimal couplings in the three models. Indeed, if we approximate the models \eqref{sqrt} and \eqref{exp} as 
$f = B \left(1 - e^{-p \sqrt{|T|}}\right) L_m \approx B p \sqrt{|T|} L_m$ and $f = B \left(1 - e^{-p T}\right) L_m \approx B p T L_m$,
their structure closely resembles that of model \eqref{BTL_m}. 

\subsection{Coupling between $\partial_\mu(f(T,L_m))$ and $J^\mu$} \label{Coupling_f}

For the CPT-violating term (\ref{General_CP}), the induced chemical potential can be written as $\mu \sim \pm\frac{\Dot{T}F_T+\partial_t(L_m)F_L}{M^2_*}$  and thus the corresponding baryon-to-entropy ratio is given by
\begin{equation}
    \frac{n_b}{s}\simeq\frac{15g_b\epsilon}{4\pi^2 g_*}\frac{\Dot{T}F_T(T,L_m)+\partial_t(L_m)F_L(T,L_m)}{M^2_*\mathbf{T}}\Bigg{|}_{\mathbf{T}_D}\, .
\end{equation}

As in previous sections, we assume that the Universe evolves gradually from one equilibrium state to another, and we adopt a power-law scale factor as in Eq.~(\ref{scale factor}). For all models, we take the matter Lagrangian to be $L_m = P = w \rho$.

\subsubsection{$f=BTL_m$\label{general_BTL_m}}

Using the energy density \eqref{BTL_m_rho} and the decoupling time \eqref{BTL_m_tD} found in \eqref{BTLm1} we find the baryon-to-entropy ratio for the CPT-violating interaction term \eqref{General_CP}
\begin{equation}
        \frac{n_b}{s}\simeq\frac{3g_b\epsilon Bwn^2\mathbf{T}_D^3}{2M_*^2}\left[t_D^{-3}+\frac{t_D^{-1}}{3n^2M_{\text{Pl}}^2}\left(\frac{\pi^2g_*\mathbf{T}_D^4}{30}\right)\right]
        	\, .
    \label{BTL_m_general_baryogeneis}
\end{equation}

\begin{figure}[t!]
    \centering
    \includegraphics[width=0.475\textwidth]{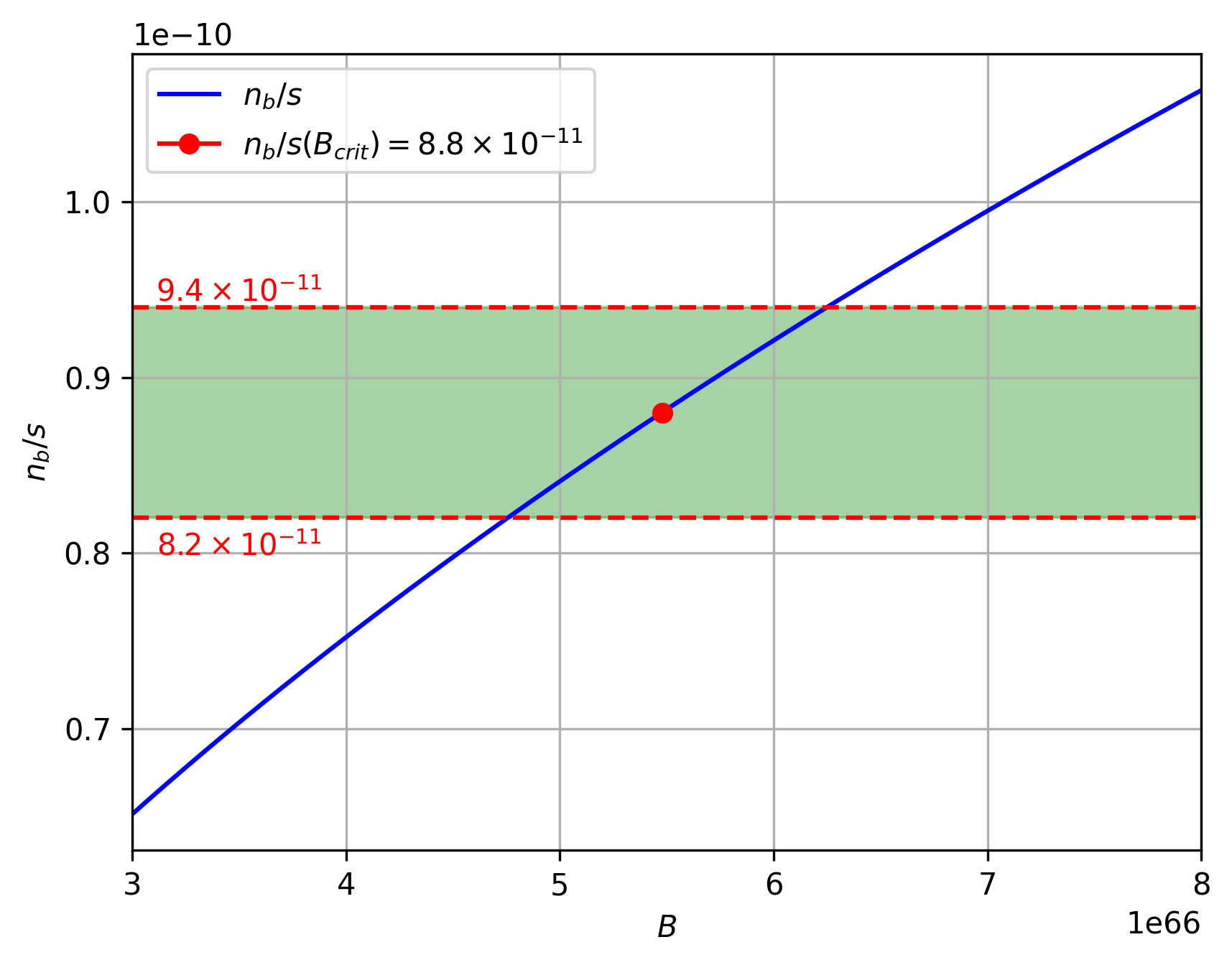}
    \caption{Plot of the baryon to entropy ratio for model $f=BTL_m$ for varying B,  $\mathbf{T}_D=10^{9}\ \text{GeV}$, $M_*=1\times 10^{13}\ \text{GeV}$, $\epsilon=1$ ,$n=\frac{1}{2}$ and $w=\frac{1}{3}$}
    \label{fig:BTLm_general}
\end{figure}

Figure~\ref{fig:BTLm_general} illustrates the dependence of the baryon-to-entropy ratio on the parameter $B$. The blue curve represents the evolution of $\frac{n_b}{s}(B)$, the dashed red lines indicate the upper and lower bounds imposed by observational constraints, and the red dot marks the critical value of $B$, with $B_{\rm crit} \approx 5.48 \times 10^{66}\,\mathrm{GeV}^{-4}$.

\subsubsection{$f=B(1-e^{-p\sqrt{|T|}})L_m$\label{general_sqrt}}

Using the energy density given in Eq.~\eqref{sqrt_rho} and the decoupling time from Eq.~\eqref{sqrt_tD}, as determined in Sec.~\eqref{sqrt1}, we compute the baryon-to-entropy ratio corresponding to the CPT-violating interaction term in Eq.~\eqref{General_CP}
\begin{equation}
    \begin{split}
        &\frac{n_b}{s}\simeq\frac{Bwp\epsilon g_b\mathbf{T}_D^3}{8  M^2_* }\left(12n^2t_D^{-3}
        \left[\left(\frac{t_D}{2\sqrt{6}n}-\frac{p}{2}\right)\right]\right.\\
        &\left.-\left(\frac{\pi^2 g_* \mathbf{T}_D^4}{30}\right)\left[\frac{2\sqrt{6}}{3nM_{\text{Pl}}^2}+Bp(1-w)t_D^{-1}\right]\right)
    \end{split}\, .
    \label{sqrt1_general_baryogenesis}
\end{equation}

\begin{figure}[t!]
    \centering
    \includegraphics[width=0.475\textwidth]{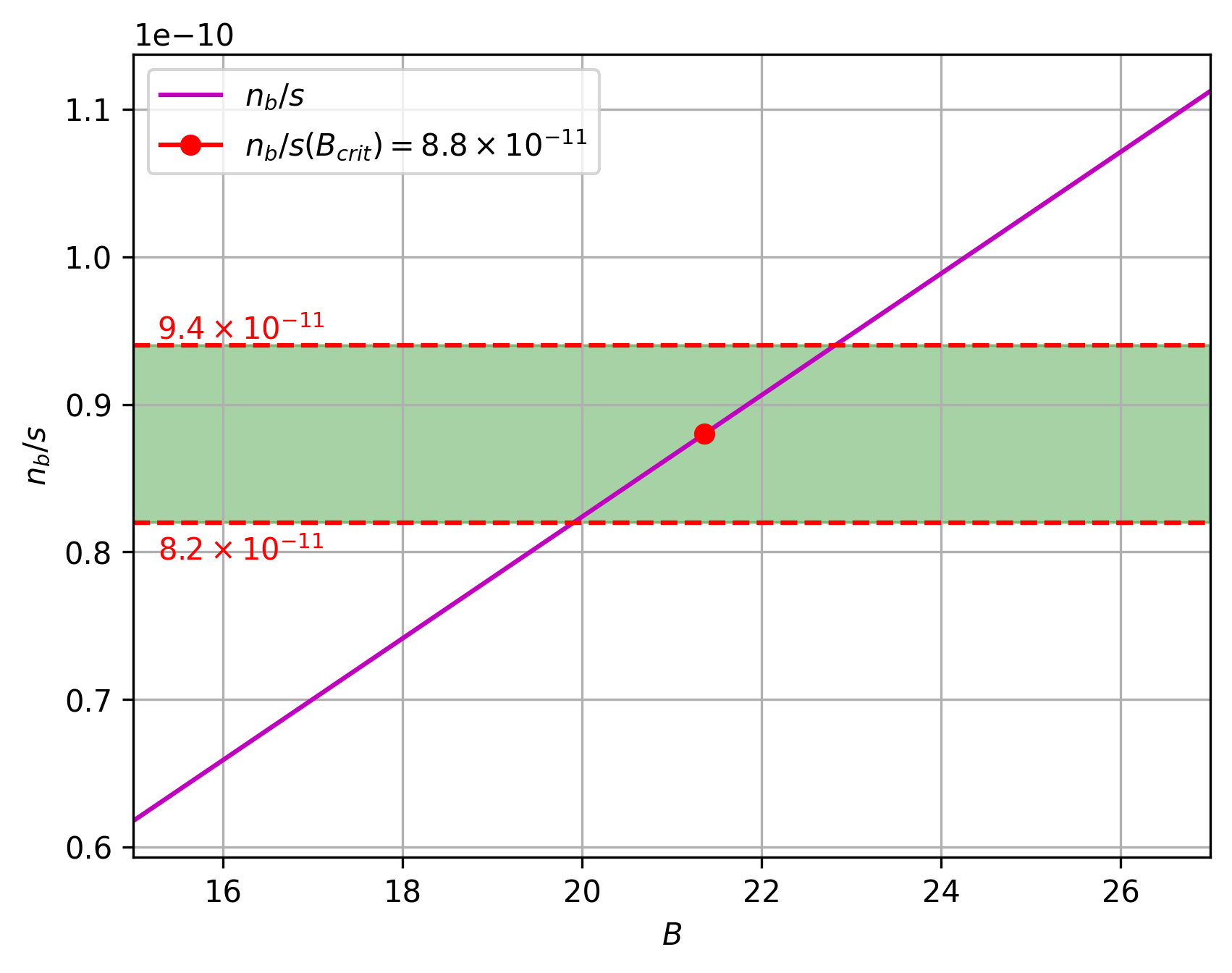}
    \caption{Plot of the baryon to entropy ratio for model $f=f=B(1-e^{-p\sqrt{T}})L_m$ for varying B,  $\mathbf{T}_D=10^{9}\ \text{GeV}$, $M_*=1\times 10^{10}\ \text{GeV}$,  $p=-10^{-11}\, \mathrm{GeV}^{-1}$, $\epsilon=-1$ , $n=\frac{1}{2}$ and $w=\frac{1}{3}$}
    \label{fig:sqrt_general1}
\end{figure}

Figure~\ref{fig:sqrt_general1} illustrates the dependence of the baryon-to-entropy ratio on the parameter $B$. The pink curve represents the evolution of $\frac{n_b}{s}(B)$, the dashed red lines indicate the upper and lower bounds imposed by observational constraints, and the red dot marks the critical value of $B$, with $B_{\rm crit} \approx 2.136 \times 10^{-6}\,\mathrm{GeV}^{-2}$.

\subsubsection{$f=B(1-e^{-pT})L_m$\label{general_exp}}

Using the energy density \eqref{exp_rho} and the decoupling time \eqref{exp_tD} found in \eqref{exp1} we find the baryon-to-entropy ratio for the CPT-Violating interaction term \eqref{General_CP}
\begin{equation}
    \begin{split}
    &\frac{n_b}{s}\simeq\frac{3\epsilon Bwpn^2 g_b \mathbf{T}_D^3}{2M_*^2}t_D^{-3}
    \Bigg[1+6pn^2t_D^{-2}	\\
    &+\left(\frac{\pi^2g_*\mathbf{T}_D^4}{30}\right)\left(\frac{t_D^2}{3n^2M_{\text{Pl}}^2}+12Bwp^2n^2t_D^{-2}\right)\Bigg]
    \end{split}
\end{equation}

\begin{figure}[t!]
    \centering
    \includegraphics[width=0.475\textwidth]{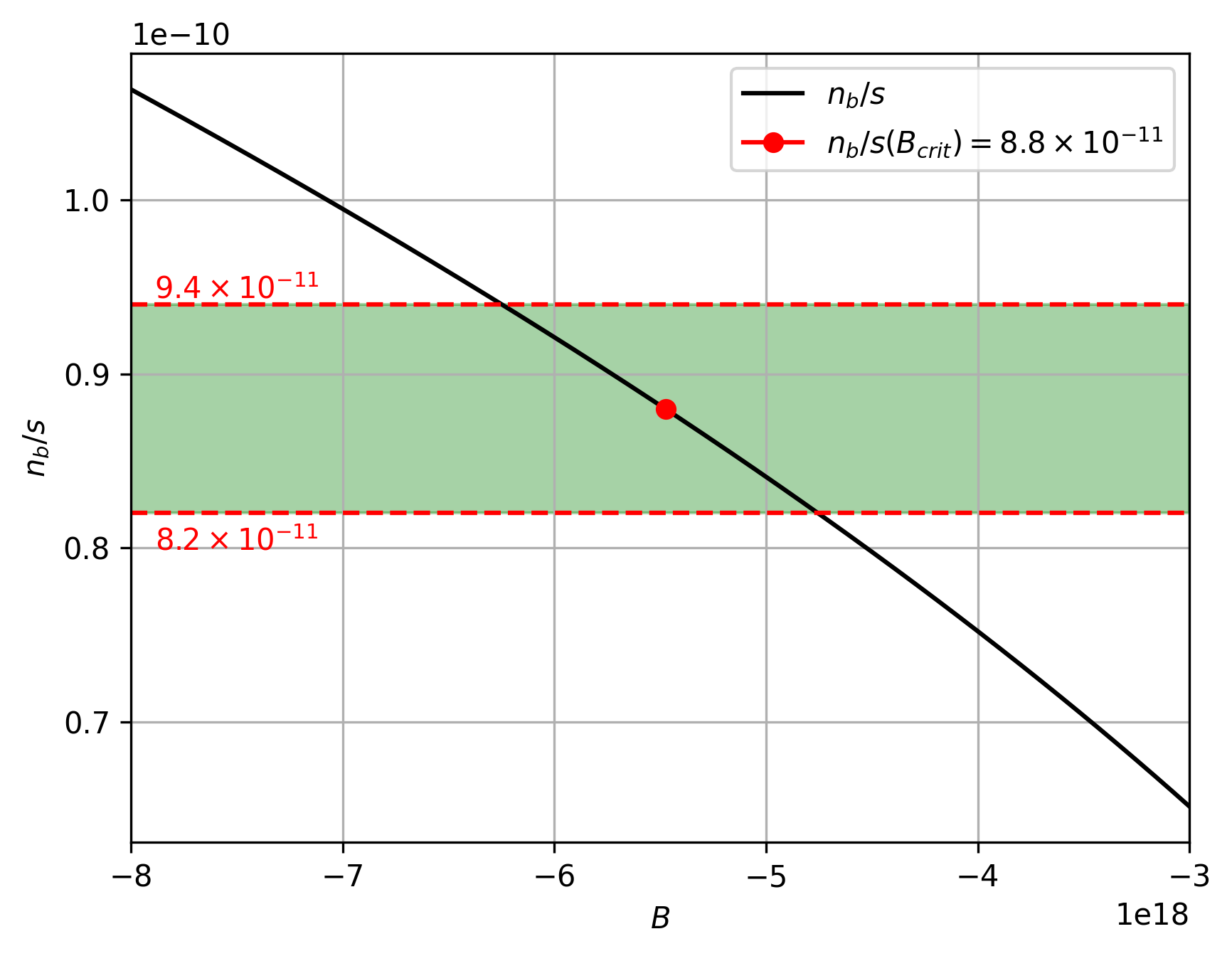}
    \caption{Plot of the baryon to entropy ratio for model $f=B(1-e^{-pT})L_m$ for varying B,  $\mathbf{T}_D=10^{9}\ \text{GeV}$, $M_*=1\times 10^{11}\ \text{GeV}$,  $p=-10^{-12}\, \mathrm{GeV}^{-2}$, $\epsilon=1$ , $n=\frac{1}{2}$ and $w=\frac{1}{3}$}
    \label{fig:exp1_general}
\end{figure}

Figure~\ref{fig:exp1_general} depicts the dependence of the baryon-to-entropy ratio on the parameter $B$. The black curve represents the evolution of $\frac{n_b}{s}(B)$, the dashed red lines indicate the upper and lower bounds imposed by observational constraints, and the red dot marks the critical value of $B$, with $B_{\rm crit} \approx -5.475 \times 10^{18}\,\mathrm{GeV}^{-2}$.

\subsubsection{Discussion}

The introduction of a general $f(T,L_m)$ coupling significantly affects the results. Analyzing now the parameters of the asymmetry, for the model $f = B T L_m$, with $\mathbf{T}_D = 10^7\,\mathrm{GeV}$ and $M_* = 1 \times 10^{10}\,\mathrm{GeV}$, we find $B_{\rm crit} \approx 5.48 \times 10^{70}\,\mathrm{GeV}^{-4}$. Maintaining the same relationship between $\mathbf{T}_D$ and $M_*$ as in Sec.~\eqref{Discussion1}, consecutive decoupling temperatures result in $B_{\rm crit}$ values differing by approximately six orders of magnitude, with smaller $B_{\rm crit}$ corresponding to higher temperatures. Comparing these results with those in Sec.~\eqref{Discussion1}, we observe a dramatic difference: for identical temperatures, $B_{\rm crit}$ differs by roughly 73 orders of magnitude. This discrepancy arises from the coupling of $F_L$ to $\partial_t(L_m)$, which introduces a $t_D^{-1}$ term that is extremely small, necessitating a much larger $B$ to compensate.

For the model $f = B \left(1 - e^{-p \sqrt{|T|}}\right) L_m$~\eqref{general_sqrt}, with $\mathbf{T}_D = 10^7\,\mathrm{GeV}$, $M_* = 10^{10}\,\mathrm{GeV}$, and $p = -10^{-11}\,\mathrm{GeV}^{-1}$, we obtain $B_{\rm crit} \approx 2.136 \times 10^{8}\,\mathrm{GeV}^{-2}$. Consecutive decoupling temperatures, with $p\sqrt{|T|}$ fixed, yield $B_{\rm crit}$ values differing by approximately five orders of magnitude. At lower temperatures, the values of $B_{\rm crit}$ differ significantly from those obtained in Sec.~\eqref{Discussion1}, but this difference decreases as the temperature increases. For instance, at $\mathbf{T}_D = 10^7\,\mathrm{GeV}$, the discrepancy in $B_{\rm crit}$ between the two couplings is 16 orders of magnitude, whereas at $\mathbf{T}_D = 10^{14}\,\mathrm{GeV}$, the difference reduces to only two orders of magnitude.

Finally, for the model $f = B \left(1 - e^{-p T}\right) L_m$~\eqref{general_exp}, with $\mathbf{T}_D = 10^7\,\mathrm{GeV}$, $M_* = 10^9\,\mathrm{GeV}$, and $p = -10^{-12}\,\mathrm{GeV}^{-2}$, we find $B_{\rm crit} \approx -5.475 \times 10^{38}\,\mathrm{GeV}^{-2}$. Maintaining the same relationship between $M_*$ and $\mathbf{T}_D$ as in Sec.~\eqref{Discussion1} and $pT$ fixed, consecutive temperatures correspond to $B_{\rm crit}$ values differing by approximately ten orders of magnitude. As in the previous model, lower temperatures yield $B_{\rm crit}$ values that differ substantially from those in Sec.~\eqref{Discussion1}, with the discrepancy decreasing at higher temperatures. Specifically, the difference is 37 orders of magnitude at $\mathbf{T}_D = 10^7\,\mathrm{GeV}$ and 13 orders of magnitude at $\mathbf{T}_D = 10^{13}\,\mathrm{GeV}$.

Once again, for all three models, with the $B$ values found, the true coupling $BM_{\text{Pl}}^2$ implies the new modified gravity terms are highly impactful, specially for lower temperatures. These values of $B$ result in $\rho_{\rm MG} \simeq -\rho_r$, yielding a very small but positive Hubble parameter for all three models. The $B$ values found in Sec.~\eqref{Discussion1} that satisfy $\left|\frac{\rho_{\rm MG}}{\rho_r}\right| \lesssim 0.10$ are insufficient to support baryogenesis, as they produce too small an asymmetry under this general coupling. Similarly, temperatures below $\mathbf{T}_D < 10^{12}\,\mathrm{GeV}$ also fail to reconcile baryogenesis with this constraint. Even relaxing the bound on $\left|\frac{\rho_{\rm MG}}{\rho_r}\right|$ does not enable successful baryogenesis for any model. This behaviour can be attributed to the fact that the general coupling requires significantly larger values of $B$ to generate the observed baryon asymmetry than those consistent with $\left|\frac{\rho_{\rm MG}}{\rho_r}\right| < 1$, which are independent of the specific form of the coupling.

\subsection{B-violation interaction}

As mentioned at the beginning, the realization of gravitational baryogenesis and by consequence gravitational baryogenesis in $f(T,L_m)$ necessitates the existence of baryon number violating interactions.

In the seminal work by Davoudiasl et al.~\cite{Davoudiasl:2004gf}, the authors proposed a mechanism for gravitational baryogenesis based on a B-violating interaction operator, denoted as $\mathcal{O}_B$, with mass dimension $D = 4 + m$, where $m > 0$. This interaction allows for the generation of a baryon asymmetry through a coupling between the derivative of a gravitational scalar (such as the Ricci scalar in GR) and the baryon current. The rate of this B-violating interaction is characterized by
\begin{eqnarray}
	\Gamma_B=\frac{\textbf{T}^{2m+1}}{M_B^{2m}}\, ,
\end{eqnarray}
where $M_B$ is the mass scale associated with $\mathcal{O}_B$. The baryon-violating processes cease at $\mathbf{T} = \mathbf{T}_D$, with $\mathbf{T}_D$ determined by $\Gamma_B= H(\mathbf{T)}$. 

In their original formulation, the interaction term induces a chemical potential for baryons, proportional to the time derivative of the Ricci scalar, $\dot{R}$, in a Friedmann–Lemaître–Robertson–Walker (FLRW) background. This chemical potential leads to a preferential production of baryons over anti-baryons in thermal equilibrium, resulting in a net baryon-to-entropy ratio. 

Considering the case of the $\partial_\mu(-T-\frac{L_m}{L_0})$ coupling~\eqref{Coupling_T_Lm} with a $D=7$ operator, the mass scale $M_B$ for each $f(T,L_m)$ model is as follows: For the first model~\eqref{BTL_m}, using the values of $B$ and $T_D$ obtained in Figure~\eqref{fig:BTL_m}, we find 
$M_B = 1.173 \times 10^{13}\ \text{GeV}$. For the second model~\eqref{sqrt}, with the values of $B$ and $T_D$ from Figure~\eqref{fig:sqrt1_normal}, we obtain 
$M_B = 1.315 \times 10^{11}\ \text{GeV}$. For the third model~\eqref{exp}, using the values of $B$ and $T_D$ shown in Figure~\eqref{fig:exp_1_normal}, we find 
$M_B = 5.447 \times 10^{11}\ \text{GeV}$.

\section{Late-time cosmology}

\begin{figure}[t!]
    \centering
    \includegraphics[width=0.475\textwidth]{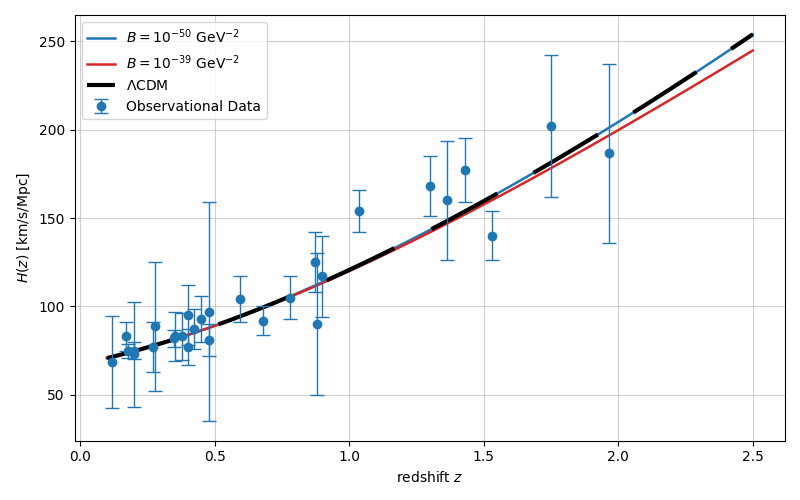}
    \caption{Evolution of the Hubble rate for the model $f = BTL_m$ with two values of the free parameter, $B = 1 \times 10^{-50}\ \text{GeV}^{-2}$ (blue) and $B = 1 \times 10^{-39}\ \text{GeV}^{-4}$ (red). The $\Lambda$CDM prediction is shown in black for comparison.}
    \label{fig:latetime_BTLm}
\end{figure}

\begin{figure}[h]
    \centering
    \includegraphics[width=0.475\textwidth]{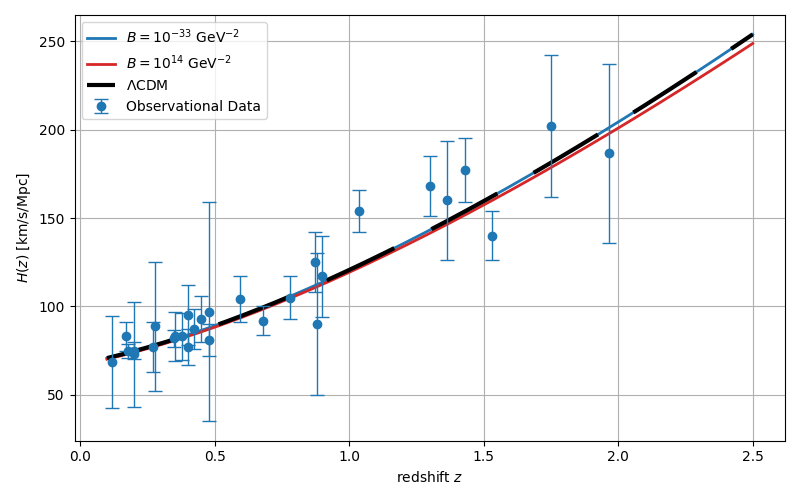}
    \caption{Evolution of the Hubble rate for the model $f = B\left(1-e^{-p\sqrt{|T|}}\right)L_m$ with two values of the free parameter, $B = 1 \times 10^{-33}\ \text{GeV}^{-2}$ (blue) and $B = 1 \times 10^{14}\ \text{GeV}^{-2}$ (red), in both cases with $p = -10^{-11}\ \text{GeV}^{-1}$. The $\Lambda$CDM prediction is shown in black for comparison.}
    \label{fig:latetime_sqrt}
\end{figure}

\begin{figure}[h]
    \centering
    \includegraphics[width=0.475\textwidth]{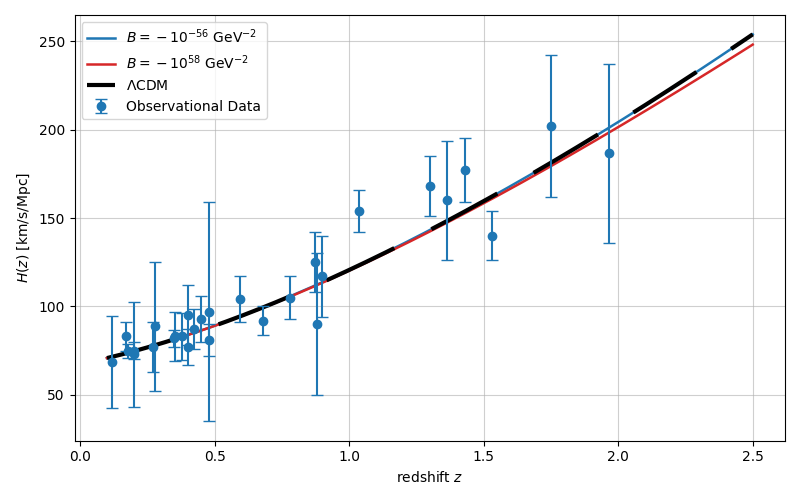}
    \caption{Evolution of the Hubble rate for the model $f = B\left(1-e^{-pT}\right)L_m$ with two values of the free parameter, $B = -1 \times 10^{-37}\ \text{GeV}^{-2}$ (blue) and $B = -1 \times 10^{58}\ \text{GeV}^{-2}$ (red), in both cases with $p = -10^{-14}\ \text{GeV}^{-2}$. The $\Lambda$CDM prediction is shown in black for comparison.}
    \label{fig:latetime_exp}
\end{figure}

Having established the constraints from gravitational baryogenesis, it is natural to further investigate the cosmological viability of the models at late times. As a preliminary consistency check, we examine their late-time evolution by plotting the Hubble parameter $H(z)$ as a function of redshift $z$. Figures~\eqref{fig:latetime_BTLm}, \eqref{fig:latetime_sqrt}, and \eqref{fig:latetime_exp} depict the evolution of the Hubble rate for each model over the redshift range $0.1 \leq z \leq 2.5$, allowing for a direct comparison with observational data~\cite{H1,H2} and with the expectations of standard cosmological behaviour.

Figures~\eqref{fig:latetime_BTLm}, \eqref{fig:latetime_sqrt}, and \eqref{fig:latetime_exp} display the Hubble rate for the corresponding models with two representative values of $B$. The blue curve corresponds to the values of $B$ determined in Sec.~\eqref{Discussion1}, while the red curve represents a value of $B$ at which the Hubble rate begins to deviate from the $\Lambda$CDM prediction. From these results, we infer that the models, when constrained by gravitational baryogenesis, remain consistent with $\Lambda$CDM in describing the late-time accelerated expansion of the Universe.

\section{Conclusion}\label{sec:conclusion}

In this paper, we have investigated the gravitational baryogenesis scenario for three models within the $f(T,L_m)$ framework using three distinct approaches. The first approach employs the interaction term given in Eq.~\eqref{eq:-T term}, exploring the effects induced by the modifications inherent to $f(T,L_m)$ gravity. The second approach extends this interaction by including the matter Lagrangian $L_m$, leading to the interaction term in Eq.~\eqref{CP-Violating}, which effectively behaves in a manner analogous to $\partial_\mu(-T)$. Finally, the third approach, referred to as general baryogenesis, considers a coupling of the baryonic current to a general $f(T,L_m)$ function, rather than being limited to the torsion scalar $T$ or its combination with $L_m$. All models analyzed in this study feature a torsion–matter coupling characterized by a free parameter $B$, with two of the three models also depending on an additional parameter $p$.

For the $\partial_\mu(-T)$ coupling~\eqref{Coupling_T}, we find that none of the three models is capable of supporting successful baryogenesis within the estimated range of decoupling temperatures, $\mathbf{T}_D \sim 10^7$--$10^{14}\,\mathrm{GeV}$. Instead, all three models require decoupling temperatures of the order of the inflationary scale, $\mathbf{T}_D \sim M_I$, which lies well beyond our considered range. The condition $\mathbf{T}_D \sim M_I$ is the same one necessary for the original gravitational baryogenesis paradigm.   

For the $\partial_\mu(-T-\frac{L_m}{L_0})$ coupling~\eqref{Coupling_T_Lm}, we find that the inclusion of the $\partial_\mu(-\frac{L_m}{L_0})$ term contributes positively to generating the baryon asymmetry, as it significantly increases the flexibility of the free parameters in the models. Nevertheless, when the modified gravity contribution is required to be a small correction to GR, as imposed by Eq.~\eqref{Constrain1}, successful baryogenesis can only be achieved for decoupling temperatures in the range $\mathbf{T}_D \sim 10^{12}$--$10^{14}\,\mathrm{GeV}$. Temperatures below this range produce an insufficient asymmetry to satisfy the observational constraints. Moreover, the estimated values of $B_{\text{crit}}$ are typically small, thereby rendering the overall coupling constant of $f(T, L_m)$ negligible compared with that associated with the Torsion-scalar term. Such results indicate that small deviations can aid in the baryogenesis mechanism.

For the $\partial_\mu(f(T,L_m))$ coupling~\eqref{Coupling_f}, we find that all three models require significantly larger values of the parameter $B$ compared to the $\partial_\mu(-T-\frac{L_m}{L_0})$ coupling, particularly at lower decoupling temperatures. Moreover when the contribution of modified gravity is constrained to be a small correction to GR, $\left|\frac{\rho_{\rm MG}}{\rho}\right| < 1$, none of the models are able to generate the observed baryon asymmetry at any temperature, as the resulting asymmetry remains insufficient to satisfy observational bounds. 

An important conceptual point is that the successful scenarios we have identified rely crucially on the non-minimal torsion--matter couplings encoded in $f(T,L_m)$. In standard GR with a radiation fluid, the Ricci scalar vanishes and the original curvature-based gravitational baryogenesis operator $\partial_\mu R\,J^\mu$ does not generate a sufficient asymmetry in the radiation era~\cite{Davoudiasl:2004gf}. By contrast, in $f(T,L_m)$ gravity the torsion scalar and, more importantly, its coupling to the matter Lagrangian $L_m$ provide a non-vanishing source for the baryon chemical potential even in the radiation-dominated regime. This allows us to obtain the observed baryon-to-entropy ratio for decoupling temperatures $10^{12}$--$10^{14}\,\mathrm{GeV}$ while keeping the modified gravity sector subdominant, a behaviour that cannot be reproduced in GR with a perfect fluid.

Furthermore, a late-time cosmological study reveals that these models present a Hubble rate evolution compatible with the standard $\Lambda$CDM scenario, reproducing the observed accelerated expansion of the Universe. These results suggest that extensions of $f(T,L_m)$ gravity, possibly including more general couplings or additional matter-sector contributions, may offer fertile ground for future investigations into the interplay between fundamental interactions and baryogenesis, as well as potential observational signatures in both the early and late Universe.

\section*{Acknowledgments}
DSP and FSNL acknowledge support from the Funda\c{c}\~{a}o para a Ci\^{e}ncia e a Tecnologia (FCT) research grants UIDB/04434/2020, UIDP/04434/2020 and PTDC/FIS-AST/0054/2021. FSNL also acknowledges support from the FCT Scientific Employment Stimulus contract with reference CEECINST/00032/2018.


\begin{thebibliography}{999}


\bibitem{Dimopoulos:1978kv}
S.~Dimopoulos and L.~Susskind,
``On the Baryon Number of the Universe,''
Phys. Rev. D \textbf{18} (1978), 4500-4509

\bibitem{wilczek1980cosmic}
Wilczek, F. The cosmic asymmetry between matter and antimatter. {\em Scientific American}. \textbf{243}, 82-91 (1980)

\bibitem{Dine:2003ax}
M.~Dine and A.~Kusenko,
``The Origin of the matter - antimatter asymmetry,''
Rev. Mod. Phys. \textbf{76} (2003), 1
[arXiv:hep-ph/0303065 [hep-ph]].

\bibitem{Cline:2006ts}
J.~M.~Cline,
``Baryogenesis,''
[arXiv:hep-ph/0609145 [hep-ph]].

\bibitem{Burles:2000ju}
S.~Burles, K.~M.~Nollett and M.~S.~Turner,
``What is the BBN prediction for the baryon density and how reliable is it?,''
Phys. Rev. D \textbf{63} (2001), 063512
[arXiv:astro-ph/0008495 [astro-ph]].

\bibitem{WMAP:2003ivt}
C.~L.~Bennett \textit{et al.} [WMAP],
``First year Wilkinson Microwave Anisotropy Probe (WMAP) observations: Preliminary maps and basic results,''
Astrophys. J. Suppl. \textbf{148} (2003), 1-27
[arXiv:astro-ph/0302207 [astro-ph]].

\bibitem{Burles:2000zk}
S.~Burles, K.~M.~Nollett and M.~S.~Turner,
``Big bang nucleosynthesis predictions for precision cosmology,''
Astrophys. J. Lett. \textbf{552} (2001), L1-L6
[arXiv:astro-ph/0010171 [astro-ph]].

\bibitem{Fields:2019pfx}
B.~D.~Fields, K.~A.~Olive, T.~H.~Yeh and C.~Young,
JCAP \textbf{03} (2020), 010
[erratum: JCAP \textbf{11} (2020), E02]
[arXiv:1912.01132 [astro-ph.CO]].

\bibitem{ParticleDataGroup:2020ssz}
P.~A.~Zyla \textit{et al.} [Particle Data Group],
PTEP \textbf{2020} (2020) no.8, 083C01.

\bibitem{WMAP:2003ogi}
C.~L.~Bennett \textit{et al.} [WMAP],
``The Microwave Anisotropy Probe (MAP) mission,''
Astrophys. J. \textbf{583} (2003), 1-23
[arXiv:astro-ph/0301158 [astro-ph]].


\bibitem{Planck:2018vyg}
N.~Aghanim \textit{et al.} [Planck],
``Planck 2018 results. VI. Cosmological parameters,''
Astron. Astrophys. \textbf{641} (2020), A6
[erratum: Astron. Astrophys. \textbf{652} (2021), C4]
[arXiv:1807.06209 [astro-ph.CO]].

\bibitem[Sakharov(1967)]{Sakharov}
A.~D.~Sakharov,
``Violation of CP Invariance, C asymmetry, and baryon asymmetry of the universe,''
Pisma Zh. Eksp. Teor. Fiz. \textbf{5} (1967), 32-35.

\bibitem{Pereira:2023xiw}
D.~S.~Pereira, J.~Ferraz, F.~S.~N.~Lobo and J.~P.~Mimoso,
``Baryogenesis: A Symmetry Breaking in the Primordial Universe Revisited,''
Symmetry \textbf{16} (2024) no.1, 13
[arXiv:2312.14080 [gr-qc]].

\bibitem{Garbrecht:2018mrp}
B.~Garbrecht,
``Why is there more matter than antimatter? Calculational methods for leptogenesis and electroweak baryogenesis,''
Prog. Part. Nucl. Phys. \textbf{110} (2020), 103727
[arXiv:1812.02651 [hep-ph]].

\bibitem{Bodeker:2020ghk}
D.~Bodeker and W.~Buchmuller,
``Baryogenesis from the weak scale to the grand unification scale,''
Rev. Mod. Phys. \textbf{93} (2021) no.3, 3
[arXiv:2009.07294 [hep-ph]].

\bibitem{Allahverdi:2012ju}
R.~Allahverdi and A.~Mazumdar,
``A mini review on Affleck-Dine baryogenesis,''
New J. Phys. \textbf{14} (2012), 125013.

\bibitem{Morrissey:2012db}
D.~E.~Morrissey and M.~J.~Ramsey-Musolf,
``Electroweak baryogenesis,''
New J. Phys. \textbf{14} (2012), 125003
[arXiv:1206.2942 [hep-ph]].

\bibitem{Riotto:1999yt}
A.~Riotto and M.~Trodden,
``Recent progress in baryogenesis,''
Ann. Rev. Nucl. Part. Sci. \textbf{49} (1999), 35-75
[arXiv:hep-ph/9901362 [hep-ph]].

\bibitem{Davoudiasl:2004gf}
H.~Davoudiasl, R.~Kitano, G.~D.~Kribs, H.~Murayama and P.~J.~Steinhardt,
``Gravitational baryogenesis,''
Phys. Rev. Lett. \textbf{93} (2004), 201301
[arXiv:hep-ph/0403019 [hep-ph]].

\bibitem{Pereira:2025flo}
D.~S.~Pereira, F.~S.~N.~Lobo and J.~Mimoso,
``Baryon asymmetry from higher-order matter contributions in gravity,''
[arXiv:2504.21504 [gr-qc]].

\bibitem{Mojahed:2024yus}
M.~A.~Mojahed, K.~Schmitz and X.~J.~Xu,
``Gravitational chargegenesis,''
[arXiv:2409.10605 [hep-ph]].

\bibitem{Bhattacharjee:2020wbh}
S.~Bhattacharjee and P.~K.~Sahoo,
``Baryogenesis in $f(Q,{\mathcal { T}})$ gravity,''
Eur. Phys. J. C \textbf{80} (2020) no.3, 289
[arXiv:2002.11483 [physics.gen-ph]].

\bibitem{Sahoo:2019pat}
P.~K.~Sahoo and S.~Bhattacharjee,
``Gravitational Baryogenesis in Non-Minimal Coupled $f(R,T)$ Gravity,''
Int. J. Theor. Phys. \textbf{59} (2020) no.5, 1451-1459
[arXiv:1907.13460 [physics.gen-ph]].

\bibitem{Li:2004hh}
H.~Li, M.~z.~Li and X.~m.~Zhang,
``Gravitational leptogenesis and neutrino mass limit,''
Phys. Rev. D \textbf{70} (2004), 047302
[arXiv:hep-ph/0403281 [hep-ph]].

\bibitem{Lambiase:2006dq}
G.~Lambiase and G.~Scarpetta,
``Baryogenesis in f(R): Theories of Gravity,''
Phys. Rev. D \textbf{74} (2006), 087504
[arXiv:astro-ph/0610367 [astro-ph]].

\bibitem{Odintsov:2016hgc}
S.~D.~Odintsov and V.~K.~Oikonomou,
``Gauss\textendash{}Bonnet gravitational baryogenesis,''
Phys. Lett. B \textbf{760} (2016), 259-262
[arXiv:1607.00545 [gr-qc]].

\bibitem{Bhattacharjee:2021jwm}
S.~Bhattacharjee,
``Baryogenesis in f(P) gravity,''
Int. J. Mod. Phys. A \textbf{36} (2021) no.27, 2150200
[arXiv:2103.15312 [gr-qc]].

\bibitem{Jaybhaye:2023lgr}
L.~V.~Jaybhaye, S.~Bhattacharjee and P.~K.~Sahoo,
``Baryogenesis in f(R,Lm) gravity,''
Phys. Dark Univ. \textbf{40} (2023), 101223
[arXiv:2304.02482 [gr-qc]].

\bibitem{Baffou:2018hpe}
E.~H.~Baffou, M.~J.~S.~Houndjo, D.~A.~Kanfon and I.~G.~Salako,
``$f(R,T)$ models applied to baryogenesis,''
Eur. Phys. J. C \textbf{79} (2019) no.2, 112
[arXiv:1808.01917 [gr-qc]].

\bibitem{Oikonomou:2016jjh}
V.~K.~Oikonomou and E.~N.~Saridakis,
``$f(T)$ gravitational baryogenesis,''
Phys. Rev. D \textbf{94} (2016) no.12, 124005
[arXiv:1607.08561 [gr-qc]].

\bibitem{Bhattacharjee:2020jfk}
S.~Bhattacharjee,
``Gravitational baryogenesis in extended teleparallel theories of gravity,''
Phys. Dark Univ. \textbf{30} (2020), 100612
[arXiv:2005.05534 [gr-qc]].

\bibitem{Mishra:2023khd}
S.~S.~Mishra, S.~Mandal and P.~K.~Sahoo,
``Constraining f(T,T) gravity with gravitational baryogenesis,''
Phys. Lett. B \textbf{842} (2023), 137959
[arXiv:2305.09707 [gr-qc]].


\bibitem{Pereira:2024ddu}
D.~S.~Pereira,
``Scalar-tensor Baryogenesis,''
[arXiv:2412.06984 [gr-qc]].

\bibitem{BeltranJimenez:2019esp}
J.~Beltr{\'a}n Jim{\'e}nez, L.~Heisenberg and T.~S.~Koivisto,
``The Geometrical Trinity of Gravity,''
Universe \textbf{5} (2019) no.7, 173
[arXiv:1903.06830 [hep-th]].

\bibitem{Heisenberg:2023lru}
L.~Heisenberg,
``Review on f(Q) gravity,''
Phys. Rept. \textbf{1066} (2024), 1-78
[arXiv:2309.15958 [gr-qc]].

\bibitem{Xu:2019sbp}
Y.~Xu, G.~Li, T.~Harko and S.~D.~Liang,
``$f(Q,T)$ gravity,''
Eur. Phys. J. C \textbf{79} (2019) no.8, 708
[arXiv:1908.04760 [gr-qc]].

\bibitem{Hazarika:2024alm}
A.~Hazarika, S.~Arora, P.~K.~Sahoo and T.~Harko,
``f(Q,Lm) gravity, and its cosmological implications,''
Phys. Dark Univ. \textbf{50} (2025), 102092
[arXiv:2407.00989 [gr-qc]].

\bibitem{Kaczmarek:2024quk}
A.~Z.~Kaczmarek, J.~L.~Rosa and D.~Szcz{\c{e}}{\'s}niak,
``Dynamical reconstruction of the $\Lambda $CDM model in the scalar{\textendash}tensor representation of $f\left( Q,T\right) $ gravity,''
Eur. Phys. J. C \textbf{85} (2025) no.2, 203
[arXiv:2410.00707 [gr-qc]].

\bibitem{Samaddar:2024qno}
A.~Samaddar and S.~S.~Singh,
``A novel approach to baryogenesis in f(Q,Lm) gravity and its cosmological implications,''
Nucl. Phys. B \textbf{1012} (2025), 116834
[arXiv:2410.05335 [gr-qc]].

\bibitem{fRLm}
T. Harko and F. S. N. Lobo, “f(R,Lm gravity,” Eur.
 Phys. J. C \textbf{70} (2010), 373-379 [arXiv:1008.4193 [gr-qc]].

\bibitem{Jaybhaye:2022gxq}
L.~V.~Jaybhaye, R.~Solanki, S.~Mandal and P.~K.~Sahoo,
``Cosmology in f(R,Lm) gravity,''
Phys. Lett. B \textbf{831} (2022), 137148
[arXiv:2205.03564 [gr-qc]].

\bibitem{Harko:2014gwa}
T.~Harko and F.~S.~N.~Lobo,
``Generalized curvature-matter couplings in modified gravity,''
Galaxies \textbf{2} (2014) no.3, 410-465
[arXiv:1407.2013 [gr-qc]].

\bibitem{Harko:2013yb}
T.~Harko, F.~S.~N.~Lobo, M.~K.~Mak and S.~V.~Sushkov,
``Modified-gravity wormholes without exotic matter,''
Phys. Rev. D \textbf{87} (2013) no.6, 067504
[arXiv:1301.6878 [gr-qc]].

\bibitem{Usman:2024cya}
M.~Usman, A.~Jawad and A.~M.~Sultan,
``Compatibility of gravitational baryogenesis in $f(Q, C)$ gravity,''
Eur. Phys. J. C \textbf{84} (2024) no.8, 868.

\bibitem{Azhar:2020coz}
N.~Azhar, A.~Jawad and S.~Rani,
``Generalized gravitational baryogenesis of well-known $f(T,TG)$ and $f(T,B)$ models,''
Phys. Dark Univ. \textbf{30} (2020), 100724
[arXiv:2009.13293 [gr-qc]].

\bibitem{Jawad:2023poj}
A.~Jawad, A.~M.~Sultan and S.~Rani,
``Viability of Baryon to Entropy Ratio in Modified Ho{\v{r}}ava{\textendash}Lifshitz Gravity,''
Symmetry \textbf{15} (2023) no.4, 824.

\bibitem{Alruwaili:2025mrc}
A.~D.~Alruwaili, N.~Azhar and A.~Jawad,
``Investigating baryon to entropy ratio phenomenon in f(R,{\ensuremath{\nabla}}R) gravity,''
Nucl. Phys. B \textbf{1018} (2025), 117067.

\bibitem{Linder}
E.~V.~Linder,
``Einstein's Other Gravity and the Acceleration of the Universe,''
Phys. Rev. D \textbf{81} (2010), 127301
[erratum: Phys. Rev. D \textbf{82} (2010), 109902]
[arXiv:1005.3039 [astro-ph.CO]].

\bibitem{exponential}
S.~Nesseris, S.~Basilakos, E.~N.~Saridakis and L.~Perivolaropoulos,
``Viable $f(T)$ models are practically indistinguishable from $\Lambda$CDM,''
Phys. Rev. D \textbf{88} (2013), 103010
[arXiv:1308.6142 [astro-ph.CO]].

\bibitem{Pereira}
R.~Aldrovandi and J.~G.~Pereira,
``Teleparallel Gravity: An Introduction,''
Fundam. Theor. Phys. 173 (2013), Springer,
ISBN 978-94-007-5142-2, 978-94-007-5143-9.

\bibitem{f(T)_and_cosmology}
Y.~F.~Cai, S.~Capozziello, M.~De Laurentis and E.~N.~Saridakis,
``f(T) teleparallel gravity and cosmology,''
Rept. Prog. Phys. \textbf{79} (2016) no.10, 106901.

\bibitem{f(TT)}
T.~Harko, F.~S.~N.~Lobo, G.~Otalora and E.~N.~Saridakis,
``$f(T,\mathcal{T})$ gravity and cosmology,''
JCAP \textbf{12} (2014), 021
[arXiv:1405.0519 [gr-qc]].

\bibitem{Harko:2014sja}
T.~Harko, F.~S.~N.~Lobo, G.~Otalora and E.~N.~Saridakis,
``Nonminimal torsion-matter coupling extension of f(T) gravity,''
Phys. Rev. D \textbf{89} (2014), 124036
[arXiv:1404.6212 [gr-qc]].


\bibitem{TheEarlyUniverse}
E.~W.~Kolb,
``The Early Universe,''
Front. Phys. \textbf{69} (1990), 1-547.

\bibitem{Pereira:2024olv}
D.~S.~Pereira, J.~P.~Mimoso and F.~S.~N.~Lobo,
``Extension of Buchdahl{\textquoteright}s Theorem on Reciprocal Solutions,''
Symmetry \textbf{16} (2024) no.7, 881
[arXiv:2407.08320 [gr-qc]].

\bibitem{Planck:2018jri}
Y.~Akrami \textit{et al.} [Planck],
``Planck 2018 results. X. Constraints on inflation,''
Astron. Astrophys. \textbf{641} (2020), A10
[arXiv:1807.06211 [astro-ph.CO]].

\bibitem{Amin:2014eta}
M.~A.~Amin, M.~P.~Hertzberg, D.~I.~Kaiser and J.~Karouby,
``Nonperturbative Dynamics Of Reheating After Inflation: A Review,''
Int. J. Mod. Phys. D \textbf{24} (2014), 1530003
[arXiv:1410.3808 [hep-ph]].

\bibitem{Bezrukov:2007ep}
F.~L.~Bezrukov and M.~Shaposhnikov,
``The Standard Model Higgs boson as the inflaton,''
Phys. Lett. B \textbf{659} (2008), 703-706
[arXiv:0710.3755 [hep-th]].

\bibitem{Bezrukov:2011gp}
F.~L.~Bezrukov and D.~S.~Gorbunov,
``Distinguishing between R$^2$-inflation and Higgs-inflation,''
Phys. Lett. B \textbf{713} (2012), 365-368
[arXiv:1111.4397 [hep-ph]].

\bibitem{Kohri:2005wn}
K.~Kohri, T.~Moroi and A.~Yotsuyanagi,
``Big-bang nucleosynthesis with unstable gravitino and upper bound on the reheating temperature,''
Phys. Rev. D \textbf{73} (2006), 123511
[arXiv:hep-ph/0507245 [hep-ph]].


\bibitem{f(T)baryogenesis}
V.~K.~Oikonomou and E.~N.~Saridakis,
``$f(T)$ gravitational baryogenesis,''
Phys. Rev. D \textbf{94} (2016) no.12, 124005
[arXiv:1607.08561 [gr-qc]].


\bibitem{H1}
M.~Moresco,
``Raising the bar: new constraints on the Hubble parameter with cosmic chronometers at z \ensuremath{\sim} 2,''
Mon. Not. Roy. Astron. Soc. \textbf{450} (2015) no.1, L16-L20
[arXiv:1503.01116 [astro-ph.CO]].

\bibitem{H2}
H.~Boumaza and K.~Nouicer,
``Growth of Matter Perturbations in the Bi-Galileons Field Model,''
Phys. Rev. D \textbf{100} (2019) no.12, 124047
[arXiv:1909.07504 [astro-ph.CO]].


\end{thebibliography}
\end{document}